%% file: TrackJet_Final.tex
\documentclass[aps,twocolumn,prd,showpacs,preprintnumbers]{revtex4-1}

\usepackage{graphicx}   
\usepackage{subfigure}
\usepackage{multirow}
\usepackage{verbatim}

\newcommand{\varz}{z}
\newcommand{\varptrel}{p_\mathrm{T}^{rel}}
\newcommand{\kt}{k_t}

\def\TeV{\ifmmode {\mathrm{\ Te\kern -0.1em V}}\else
                   \textrm{Te\kern -0.1em V}\fi}%
\def\GeV{\ifmmode {\mathrm{\ Ge\kern -0.1em V}}\else
                   \textrm{Ge\kern -0.1em V}\fi}%
\def\MeV{\ifmmode {\mathrm{\ Me\kern -0.1em V}}\else
                   \textrm{Me\kern -0.1em V}\fi}%
\usepackage[breaklinks,colorlinks,linkcolor=blue,citecolor=blue,filecolor=blue,urlcolor=blue]{hyperref}

\begin{document}

\begin{minipage}[t]{\textwidth}
  \begin{flushright}
    CERN-PH-EP-2011-110\\
    Submitted to Physical Review D
    \end{flushright}
\end{minipage}
\vspace{0.5cm}

\title{Properties of jets measured from tracks in proton-proton collisions at center-of-mass energy $\sqrt{s}=7\,\TeV$ with the ATLAS detector}

\author{The ATLAS Collaboration}

\date{August 4, 2011}

\begin{abstract}
Jets are identified and their properties studied in center-of-mass energy $\sqrt{s} = 7\,\TeV$ proton-proton collisions at the Large Hadron Collider 
using charged particles measured by the ATLAS inner detector.
Events are selected using a minimum bias trigger,
allowing jets at very low transverse momentum to be observed 
and their characteristics in the transition to high-momentum fully perturbative jets to be studied.
Jets are reconstructed using the anti-$\kt$ algorithm applied to charged particles  with
two radius parameter choices, 0.4 and 0.6.
An inclusive charged jet transverse momentum
cross section measurement from $4\,\GeV$ to $100\,\GeV$ is shown for four ranges in rapidity extending to 1.9
 and corrected to charged particle-level truth jets.
The transverse momenta and longitudinal momentum fractions of charged particles
within jets are measured, along with the charged particle multiplicity and the particle density
as a function of radial distance from the jet axis. 
Comparison of the data with the theoretical models implemented in existing tunings of
Monte Carlo event generators indicates reasonable overall agreement between data and Monte Carlo.
These comparisons are sensitive to Monte Carlo parton showering, hadronization, and soft physics models.
\end{abstract}


\maketitle

\section{Introduction}

Quantum Chromodynamics (QCD)~\cite{Gross:1973id,Politzer:1973fx} provides an excellent description of the kinematic distribution of 
high transverse momentum jets in proton-proton collisions, but does not give straightforward predictions for the properties of particles
within these jets or for the properties of low-momentum jets. 
These quantities may be predicted by Monte Carlo event generators, whose results are dependent on phenomenological models of
parton showering, hadronization, and soft (i.e. low momentum transfer) physics.  
These models have free parameters that must in turn be tuned to data.

In this work the properties of low-momentum jets are measured from charged particle tracks 
in the ATLAS 2010 data at center-of-mass energy $\sqrt{s} = 7\,\TeV$.
These properties will be compared to a range of Monte Carlo tunes derived from previously measured data,
allowing the study of the transition between
the MC generators' separately-implemented models of soft strong interactions and perturbative QCD.
This measurement is complementary to others that are used for comparison with models of soft QCD,
such as inclusive charged particle kinematic distributions~\cite{:2010ir}
and underlying event properties~\cite{Aad:2011qe}, and may shed light on the
limited ability of existing models to describe both types of measurements simultaneously with a single set of tunable parameters~\cite{ATL-PHYS-PUB-2011-008}.
At higher momenta, track-based jet measurements provide a complement to calorimeter-based
measurements~\cite{Aad:2011kq,daCosta:2011ni}, with results independent of calorimeter calibrations, selections, and uncertainties.

Jets have previously been measured using charged particles in proton-proton collisions at the CERN Intersecting Storage Rings~\cite{Clark:1979vc},
and in proton-antiproton collisions at the CERN Super Proton Synchrotron~\cite{Arnison:1982fh} and Fermilab Tevatron~\cite{Affolder:2001xt}.
The momentum fraction of charged particles with respect to calorimeter jets has also been measured~\cite{Arnison:1983gv,Acosta:2002gg}.
Related fragmentation measurements in electron-positron and electron-proton collisions are reviewed in Ref.~\cite{Nakamura:2010zzi}. 

In this work, jets are reconstructed using the anti-$\kt$ algorithm~\cite{antikt},
using a radius parameter $R$ of 0.4 or 0.6, applied to measured charged particles with transverse momentum $p_\mathrm{T} > 300\,\MeV$;
jet four-momenta are determined by adding constituent four-vectors.
The distribution of jet momenta, jet track multiplicity, and kinematic properties of tracks within these track-based
jets are corrected back to truth-level \emph{charged particle jets}, which are defined
to be the jets obtained when the same algorithm is applied to all primary charged particles emerging from the proton-proton collision with the same $p_\mathrm{T}$ cut.

Five quantities are measured for charged particle jets:

\begin{equation} \frac{d^2\sigma_{jet}}{dp_{\mathrm{T}jet}dy_{jet}} ;\; \frac{1}{N_{jet}}\frac{dN_{jet}}{dN^{ch}_{jet}} ;\; \frac{1}{N_{jet}}\frac{dN_{ch}}{d\varz} ;\;
\frac{1}{N_{jet}}\frac{dN_{ch}}{d\varptrel} ;\; \rho_{ch}(r) \end{equation}

\noindent The inclusive charged particle jet cross section $\sigma_{jet}$ is measured in bins of $p_{\mathrm{T}jet}$ for four rapidity ($y$)~\footnote{
ATLAS uses a right-handed coordinate system with its origin at the nominal interaction point (IP) in the center 
of the detector and the Z-axis along the beam pipe. The X-axis points from the IP to the center of the LHC ring, 
and the Y axis points upward. Cylindrical coordinates $(r,\phi)$ are used in the transverse plane, 
$\phi$ being the azimuthal angle around the beam pipe. 
The rapidity $y$ for a track or jet is defined by $y = 0.5 \times \ln [(E+p_\mathrm{Z})/(E-p_\mathrm{Z})]$,
where $E$ denotes the energy and $p_\mathrm{Z}$ is the component of the momentum along the beam direction;
the pion mass is assumed for reconstructed tracks.
}
ranges: 
0--0.5, 0.5--1.0, 1.0--1.5, 1.5--1.9,
where the last range is narrower than the others to ensure that measured jets are fully contained in the tracking acceptance of the ATLAS detector.  
The distribution of charged particle multiplicities per jet $N^{ch}_{jet}$ is measured for each of these ranges, separated further into five jet transverse momentum ranges: 
4--6$\,\GeV$, 6--10$\,\GeV$, 10--15$\,\GeV$, 15--24$\,\GeV$, 24--40$\,\GeV$.
The longitudinal momentum fraction $\varz$ of charged particles in these jets is measured in the same rapidity and transverse momentum ranges.
The variable $\varz$, also known as the fragmentation variable, is defined for each particle in a jet by:

\begin{equation} \varz = \frac{\vec{p}_{ch} \cdot \vec{p}_{jet}}{|\vec{p}_{jet}|^2}, \end{equation}

\noindent where $\vec{p}_{ch}$ is the momentum of the charged particle and $\vec{p}_{jet}$ is the momentum of the jet that contains it.
The $\varz$ distribution presented here differs from the usual definition (in~\cite{Arnison:1983gv}, for example), 
which would include neutral particles and low-momentum charged particles in the total jet momentum.
The variable $\varptrel$ is the momentum of charged particles in a jet transverse to that jet's axis:

\begin{equation} \varptrel = \frac{|\vec{p}_{ch} \times \vec{p}_{jet}|}{|\vec{p}_{jet}|}. \end{equation}

\noindent Finally, the density of charged particles in $\phi$-$y$ space, $\rho_{ch}(r)$, 
is measured as a function of the radial distance $r$ of charged particles from the axis of the jet that contains them, where $r$ is given by

\begin{equation} \label{eqn:rdef} r  = \sqrt{(\phi_{ch} - \phi_{jet})^2+(y_{ch} - y_{jet})^2}, \end{equation}

\noindent so that

\begin{equation}  \rho_{ch}(r) = \frac{1}{N_{jet}} \frac{dN_{ch}}{2 \pi r dr}. \end{equation}


\noindent Note that this is a particle number density, rather than the related energy density variable used for calorimeter-based jet shape measurements.  
The variables $\rho_{ch}(r)$ and $\varptrel$ are measured in the same rapidity and jet momentum ranges as $\varz$.

\section{The ATLAS Detector}

The ATLAS detector~\cite{Aad:2008zzm} at the Large Hadron Collider (LHC)~\cite{Evans:2008zzb} covers almost the whole solid angle
around the collision point with layers of tracking detectors, calorimeters and muon chambers. It has been
designed to study a wide range of physics topics at LHC energies. For the measurements presented in
this paper, the tracking devices and the trigger system are used.

The ATLAS inner detector (ID) has full coverage in the pseudorapidity range $|\eta| < 2.5$.
It consists of a silicon pixel detector (Pixel), a silicon microstrip detector (SCT) and a transition radiation tracker (TRT),
the last of which only covers $|\eta| < 2.0$. 
These detectors cover a sensitive radial distance from the interaction point of 50--150$\,\mathrm{mm}$, 
299--560$\,\mathrm{mm}$ and 563--1066$\,\mathrm{mm}$, respectively, and are immersed in a $2\,\mathrm{T}$ axial magnetic field.
The inner-detector barrel (end-cap) parts consist of 3 (2$\times$3) Pixel layers, 4 (2$\times$9) double-layers of single-sided silicon microstrips with a 40 mrad stereo angle, and 73 
(2$\times$160) layers of TRT straws. Typical
 position resolutions are 10, 17 and 130$\,\mathrm{\mu m}$ for the $R$-$\phi$ coordinate of the pixel detector, SCT, and TRT respectively; 
the resolution of the second measured coordinate is 115 (580)$\,\mathrm{\mu m}$ for the pixel detector (SCT). 
A track from a charged particle traversing the
barrel detector would typically have 11 silicon hits (3 pixel clusters and 8 strip clusters) and more than
30 straw hits.

The ATLAS detector has a three-level trigger system: Level 1 (L1), Level 2 (L2) and Event Filter (EF). 
The L1 triggers use custom fast electronics; the L2 and EF further refine selections in software, with
L2 using a subset of the event information for rapid processing and EF reconstructing complete events. 
For this measurement, the trigger relies on the L1 signals from the Beam Pickup Timing devices 
(BPTX) and the Minimum Bias Trigger Scintillators (MBTS). The BPTX stations are composed
of electrostatic button pick-up detectors attached to the beam pipe at $\pm 175\,\mathrm{m}$ from the center of the
ATLAS detector. The coincidence of the BPTX signal between the two sides of the detector is used to
determine when bunches are colliding in the center of the detector. The MBTS are mounted on each side of the detector 
at Z = $\pm 3.56\,\mathrm{m}$. They
are segmented into eight sectors in azimuth and two rings in pseudorapidity ($2.09 < |\eta| < 2.82$ and $2.82 < |\eta| < 3.84$). 
Data were collected for this analysis using a trigger requiring a BPTX coincidence
and MBTS trigger signals. The MBTS trigger used for this paper is configured
to require one hit above
threshold from either side of the detector, referred to as a single-arm trigger. 

\section{Monte Carlo samples} \label{mc}

The corrections used in this analysis, from detector-level track jets to truth-level charged particle jets,
are derived from a Monte Carlo (MC) simulated sample using the Pythia 6.421 event generator program~\cite{pythia} with the ATLAS AMBT1 tune,  
whose parameters are chosen based on single track distributions in ATLAS minimum bias data~\cite{:2010ir}.  
In order to derive corrections, these events are then passed through the ATLAS detector simulation~\cite{siminf}, based on GEANT~4~\cite{geant}.
Large simulated samples of the ATLAS MC09~\cite{ATL-PHYS-PUB-2010-002} and Perugia 2010~\cite{Skands:2010ak} tunes of Pythia 6 are also used.

Additional Pythia 6.421 tunings are used at the generator-level for comparison to data,
and in order to explore the sensitivity of the measurement
to underlying event, fragmentation, and hadronization parameters, as described in Section~\ref{syst}.
The Perugia family of tunes~\cite{Skands:2010ak} uses the CTEQ5L~\cite{Lai:1999wy} parton distributions;
there is a central value (``Perugia 0'') and several variants that attempt to bracket the possible changes in tune parameters that are permitted 
in fitting existing data.
Two variants, ``Perugia HARD'' and ``Perugia SOFT'', are created by changing the initial-state radiation cutoff scale so that the perturbative contribution is different,
while other parameters are adjusted to bring the tuned distributions back into agreement.  
Perugia HARD (SOFT) has more (fewer) jets per event with higher (lower) average momentum, and a higher (lower) average charged multiplicity.
The Perugia 2010 variant is an adjustment of Perugia 0 that improves the description of jet shapes and adjusts hadronization parameters
for better consistency with LEP data, given that the original values for these parameters were based on $Q^2$ shower ordering;
in particular, it has an adjusted yield of strange particles.
Perugia NOCR (``no color reconnection'') represents an effort to reproduce the same data with no color reconnection model used;
it is used in computing systematic uncertainties, but because it could not be made to fit all the original tuning data it is not included on comparison plots.
All Pythia tunes described thus far use $p_\mathrm{T}$-ordered showering; the DW tune is also used in order to include the impact of $Q^2$ ordering.

Predictions from separately-implemented generators are used for additional studies and comparisons. 
Phojet 1.12.1.3~\cite{Engel:1995yda}, which relies on Pythia 6.115 for the fragmentation of partons, is used with its default settings.
A Herwig++ 2.4.2~\cite{Bahr:2008pv} sample is used with its default settings,
along with an additional sample tuned on $7\,\TeV$ underlying event data with Herwig++ 2.5.1 (UE7)~\cite{Gieseke:2011na}.
A Pythia 8.145~\cite{Sjostrand:2007gs} sample is used with the 4C tune~\cite{Corke:2010yf}.  

Single- and double-diffractive events are included in the Phojet and Pythia 8 samples, with cross sections 
relative to the non-diffractive events as indicated by those generators.
Herwig++ does not include diffractive modeling, while Pythia 6 diffractive modeling produces a negligible yield of jets as defined in this analysis.

\section{Data selection}

This measurement uses a sample of early ATLAS data at center-of-mass energy $\sqrt{s} = 7\,\TeV$ for which the minimum bias trigger was minimally prescaled, 
corresponding to a total integrated luminosity of 799 $\mathrm{\mu b^{-1}}$ and a peak luminosity of $6.6 \times 10^{28}\: \mathrm{cm^{-2} s^{-1}}$.
Events from colliding proton bunches are selected if the MBTS recorded one or more counters above threshold on either side.  
They are further required to have a primary vertex~\cite{Piacquadio:2008zzb} reconstructed using beam-spot information~\cite{beamspot}.
Events with additional reconstructed primary vertices are rejected.  
The average number of collisions per bunch crossing $\mu$ depends on luminosity;
the highest-luminosity data collected have $\mu\sim0.14$, but over half the data have $\mu\lesssim0.01$.

\subsection{Track reconstruction} \label{tracksel}

Tracks are reconstructed using the ATLAS primary silicon-seeded tracking algorithm, with a configuration similar to that described in the $900\,\GeV$ and 
$7\,\TeV$ minimum bias measurements~\cite{:2010ir,Aad:2010rd}.
In order to select good-quality tracks emerging from the primary vertex while maintaining a high efficiency, each track is then required to have:

\begin{list}{$\bullet$}{
    \setlength{\itemsep}{0pt}
     \setlength{\parsep}{3pt}
     \setlength{\topsep}{3pt}
     \setlength{\partopsep}{0pt}
  }

\item transverse momentum $p_\mathrm{T} > 300\,\MeV$;

\item pseudorapidity $|\eta| < 2.5$;

\item transverse impact parameter with respect to the primary vertex $|d_\mathrm{0}| < 1.5\,\mathrm{mm}$ (0.2$\,\mathrm{mm}$) for tracks with $p_\mathrm{T}$ less than (greater than) $10\,\GeV$;

\item longitudinal impact parameter with respect to the primary vertex $z_\mathrm{0}$ satisfying $|z_\mathrm{0}\mathrm{sin}\theta| < 1.5\,\mathrm{mm}$;

\item if a signal (or \emph{hit}) is expected in the innermost pixel detector layer 
(i.e. if the extrapolated track passes through a section of that layer with functioning instrumentation),
 then such a hit is required, with one pixel hit in any layer required otherwise;

\item at least 6 SCT hits.

\end{list}

\subsection{Selection efficiencies} \label{seleff}

The efficiency for triggering given the presence of one or more jets is determined from data
using a random subset of all events with a BPTX coincidence.
In such events for which an offline track jet (as defined in Section~\ref{jetdef}) is present,
the fraction of events for which the MBTS single-arm trigger is also passed is determined.
It is found to be negligibly different from 100$\%$ for all events containing jets as defined in this analysis.

The efficiency for primary vertex-finding given the presence of one or more
jets is determined from data by removing the track selection cuts
that use impact parameter with respect to the primary vertex described in Section \ref{tracksel}.
Other cuts are kept as described, and a requirement is added that
the transverse impact parameter with respect to the beam spot satisfies $|d_\mathrm{0}| < 4.0\,\mathrm{mm}$.
Jets are then reconstructed from tracks satisfying this new selection,
and the fraction of events containing jets that also have a primary vertex
reconstructed is determined.   It is found to be negligibly different from 100$\%$ for all events containing jets as defined in this analysis.

The inefficiency introduced by the requirement that no additional primary vertices be present arises from two causes:

\begin{list}{$\bullet$}{
    \setlength{\itemsep}{0pt}
     \setlength{\parsep}{3pt}
     \setlength{\topsep}{3pt}
     \setlength{\partopsep}{0pt}
  }

\item Events may be lost due to multiple collisions in a single bunch-crossing.  The rate at which this occurs is calculated as a function of $\mu$~\cite{Aad:2011dr},
and corresponds to a 3.3$\%$ correction to the cross section over the entire dataset. 

\item Events may be lost because multiple vertices are incorrectly reconstructed in bunch crossings with only a single collision.
The rate at which this occurs is determined from simulated events, and parameterized as a function of the number of selected tracks in the event.
It corresponds to a cross section correction of roughly 2$\%$, depending weakly on jet momentum.

\end{list}

The efficiencies for track reconstruction and selection are derived from simulated events, whose tracking properties have been shown to be in excellent agreement with the data~\cite{:2010ir}.
The uncertainties on these efficiencies, and their impact on the present results, are discussed in Section~\ref{syst}.

\section{Jet definition} \label{jetdef}

Reconstructed track jets are produced by applying the anti-$\kt$ algorithm with radius parameter $R$ = 0.4 or 0.6 to the selected tracks;
the pion mass is assumed for all tracks in the application of the algorithm. 
After the algorithm is applied, track jets with $p_\mathrm{T} > 4\,\GeV$ and $|y| < 1.9$ are accepted, including those with only one constituent track.
The rapidity cut is chosen to ensure that jets are fully contained within the tracking acceptance $|\eta| < 2.5$ in the $R$ = 0.6 case;
for the narrower $R$ = 0.4 jets, the same rapidity cut is used for consistency.

Truth-level charged particle jets similarly have the anti-$\kt$ algorithm applied to primary charged simulated particles with $p_\mathrm{T} > 300\,\MeV$,
using each particle's true mass in the application of the algorithm.  Primary charged particles are
defined as charged particles with a mean lifetime  $\tau > 0.3 \times 10^{-10}\,\mathrm{s}$, 
which are produced in the primary collision or from subsequent decays of particles with a shorter lifetime.
Thus the charged decay products of $K^0_S$ particles are not included.
Charged particle jets are required to meet the $p_\mathrm{T}$ and $y$ requirements given for track jets above.

\section{Correction procedure} \label{analysis}

After all distributions are measured for reconstruction-level track jets, each distribution is corrected to truth-level charged particle jets.
The corrections are derived from the AMBT1 sample discussed in Section~\ref{mc}, and account for tracking efficiency and momentum resolution.

Reconstructed jets are binned simultaneously in $p_\mathrm{T}$, 
constituent multiplicity ($N^{ch}_{jet}$), and rapidity ($y$), so that a three-dimensional distribution is produced
for the purpose of applying corrections.
Similarly, for each per-track variable $\varz$, $p_{\mathrm{T}}^{rel}$, and $r$,
each track in each jet is binned in the variable, the parent jet momentum, and parent jet rapidity.
For both the jet-level quantities and each track-level variable, corrections are applied simultaneously in the three binned variables
using the Bayesian Iterative Unfolding algorithm~\cite{D'Agostini:1994zf}.

The corrections for this algorithm are based on a \emph{response matrix} derived from MC events, which encapsulates the probability
for a charged particle jet with a particular ($p_\mathrm{T}$,$N^{ch}_{jet}$,$y$) to be reconstructed in each possible $p_\mathrm{T}$, $N^{ch}_{jet}$, and $y$ bin.
A reconstructed track jet is defined to be matched to a truth-level charged particle jet if it is within
$\Delta R < 0.2$ (0.3) for jet radius 0.4 (0.6).
These values are chosen to avoid matching ambiguity while still giving good efficiency, which rises from roughly $55\%$ at the very lowest momenta to greater than $95\%$
for $p_{\mathrm{T},jet}>10\,\GeV$.
In the case of the $\varz$ response matrix, binned in ($\varz$,$p_{\mathrm{T}jet}$,$y_{jet}$), 
if a truth jet is unmatched then all its tracks are counted as lost to inefficiency.
If it is matched, then each of its constituent charged particles are matched to the closest track within $\Delta R < 0.04$; 
if there is no such track, then that particular particle is counted as lost to inefficiency.
The $\varptrel$ and $r$ matrices are filled in the same manner as the $\varz$.

The Bayesian Iterative Unfolding algorithm takes both inefficiencies and resolution effects into account, as ``missed'' entries and bin-to-bin transfers respectively;
the probability of these occurrences is determined from the central AMBT1 MC sample.
Before the Bayesian Iterative Unfolding algorithm can be applied, a correction must be made for unmatched reconstructed jets,
and for unmatched reconstructed tracks in the $\varz$, $p_{\mathrm{T}}^{rel}$, and $r$ distributions.
This correction is determined from the central MC sample by the ratio between the reconstructed distributions and the matched reconstructed distributions.

Three iterations of the algorithm are used. This is validated by applying the unfolding to ensembles of data-sized MC samples, and observing the number of iterations required to obtain convergence.
Statistical uncertainties on the unfolded variables are determined by a toy histogram-variation method,
which produces 100 pseudoexperiment distributions by varying the contents of each input bin randomly, then calculating the covariance matrix from the variation in the output of the unfolding.
This error calculation is also validated using ensembles of data-sized MC samples (with full detector simulation) by considering the distribution of pulls:

\begin{equation} \mathrm{Pull} = \frac{N_{corr} - N_{truth}}{\sigma_{corr}} \end{equation}

\noindent where $N_{corr}$ is the corrected number of events in a given bin, $N_{truth}$ is the true number of events in that bin, and
$\sigma_{corr}$ is the statistical error for that bin as reported by the unfolding procedure.  The RMS of the pull distribution should be consistent with one. In some bins it is not, so the statistical errors are scaled appropriately.  
This scaling is at most roughly 1.5 except for a few outlying bins in the multiplicity and $p_\mathrm{T}^{rel}$ distributions that have very low statistics.

The ensemble studies also reveal a bias in the unfolding algorithm calculation in closure tests, which is roughly $2\%$ in the jet $p_\mathrm{T}$ distributions.
The possibility of an additional bias from changing the prior distribution used as input to the unfolding algorithm was investigated. After three iterations, this bias vanishes.

\section{Systematic Uncertainties} \label{syst}

Systematic uncertainties are computed by determining the impact of a given effect on reconstruction-level distributions,
then propagating it through the unfolding procedure to determine the final impact on the measurement.
A summary of all systematic uncertainties for selected bins is shown in Table~\ref{table-systsummary-GeV} for the cross section and $\varptrel$ distributions, 
and in Table~\ref{table-systsummary-unitless} for the multiplicity, $\varz$, and $\rho(r)$ distributions.
The uncertainties given in each column of the tables are: 

\begin{table*}[htb!]
\begin{center}

\caption{ Summary of systematic uncertainties for selected bins in selected cross section and $\varptrel$ distributions, for $R$ = 0.6.
An overall luminosity uncertainty of $3.4\%$ is not shown, and applies to the cross section only.
\label{table-systsummary-GeV}}
\vspace { 2mm }
\begin{tabular}{| c | c | c | c | c | c | c | c | c | c |}
\hline
Distribution & Bin & Track & Track & Monte & High-$p_\mathrm{T}$ & Unmatched & Split & Clo- & Total   \\
 & [GeV] & eff. & res. & Carlo & Tracks  & Jets/Tracks & Vertex & sure &  \\
\hline \multirow{2}{*}{$\frac{d^2\sigma_{jet}}{dp_{\mathrm{T}jet}dy_{jet}}$}
& \footnotesize{4 -- 5}  &  ${}_{-3.3\%}^{+3.2\%}$ & $\pm 0.07\%$ & ${}_{-1.2\%}^{+0.07\%}$ 
& $<0.005\%$ & $\pm 0.21\%$ & $\pm 1.8\%$ & $\pm 2.0\%$ & ${}_{-4.4\%}^{+4.2\%}$ \\
\cline{2-10}
& \footnotesize{20 -- 22}  &  ${}_{-6.3\%}^{+6.6\%}$ & $\pm 0.34\%$ & ${}_{-0.69\%}^{+2.2\%}$ 
& ${}_{-0.14\%}^{+0.10\%}$ & $\pm 0.05\%$ & $\pm 2.0\%$ & $\pm 0.54\%$ & ${}_{-6.7\%}^{+7.3\%}$ \\
\cline{2-10} $\scriptstyle |y_{jet}| < 0.5$
&  \footnotesize{40 -- 45}  &  ${}_{-6.9\%}^{+6.8\%}$ & $\pm 0.39\%$ & ${}_{-2.3\%}^{+1.4\%}$ 
& ${}_{-1.4\%}^{+1.1\%}$ & $\pm 0.01\%$ & $\pm 2.1\%$ & $\pm 0.50\%$ & ${}_{-7.7\%}^{+7.3\%}$ \\
\cline{2-10}
& \footnotesize{80 -- 90}  &  ${}_{-6.9\%}^{+7.1\%}$ & $\pm 0.34\%$ & ${}_{-7.5\%}^{+7.1\%}$ 
& ${}_{-9.1\%}^{+5.5\%}$ & $<0.005\%$ & $\pm 2.3\%$ & $\pm 0.78\%$ & ${}_{-14\%}^{+12\%}$ \\
\hline \multirow{2}{*}{$\frac{d^2\sigma_{jet}}{dp_{\mathrm{T}jet}dy_{jet}}$}
& \footnotesize{4 -- 5}  &  ${}_{-5.3\%}^{+5.4\%}$ & $\pm 0.02\%$ & ${}_{-2.1\%}^{+3.0\%}$ 
& $<0.005\%$ & $\pm 0.19\%$ & $\pm 1.8\%$ & $\pm 5.2\%$ & ${}_{-7.9\%}^{+8.3\%}$ \\
\cline{2-10}
& \footnotesize{20 -- 22}  &  ${}_{-9.5\%}^{+10\%}$ & $\pm 0.02\%$ & ${}_{-3.2\%}^{+2.6\%}$ 
& ${}_{-0.15\%}^{+0.10\%}$ & $\pm 0.09\%$ & $\pm 2.0\%$ & $\pm 0.96\%$ & ${}_{-10\%}^{+11\%}$ \\
\cline{2-10} $\scriptstyle 1.5 < |y_{jet}| < 1.9$
& \footnotesize{40 -- 45}  &  ${}_{-10\%}^{+11\%}$ & $\pm 0.30\%$ & ${}_{-1.5\%}^{+2.9\%}$ 
& ${}_{-1.3\%}^{+1.1\%}$ & $\pm 0.07\%$ & $\pm 2.1\%$ & $\pm 0.95\%$ & $\pm 11\%$ \\
\cline{2-10}
& \footnotesize{80 -- 90}  &  ${}_{-11\%}^{+12\%}$ & $\pm 1.3\%$ & ${}_{-12\%}^{+7.3\%}$ 
& ${}_{-8.2\%}^{+7.3\%}$ & $\pm 9.3\%$ & $\pm 2.0\%$ & $\pm 6.9\%$ & ${}_{-21\%}^{+20\%}$ \\
\hline $\frac{1}{N_{jet}}\frac{dN_{ch}}{d\varptrel}$
& \footnotesize{0.0 -- 0.05}  &  ${}_{-0.32\%}^{+0.44\%}$ & $\pm 0.18\%$ & ${}_{-4.8\%}^{+16\%}$ 
& $<0.005\%$ & $\pm 2.7\%$ & $<0.005\%$ & $\pm 0.17\%$ & ${}_{-5.6\%}^{+16\%}$ \\
\cline{2-10} $\scriptstyle |y_{jet}| < 1.9$
& \footnotesize{0.5 -- 0.55}  &  ${}_{-0.20\%}^{+0.30\%}$ & $<0.005\%$ & ${}_{-1.7\%}^{+3.0\%}$ 
& $<0.005\%$ & $\pm 3.2\%$ & $\pm 0.01\%$ & $\pm 0.09\%$ & ${}_{-3.6\%}^{+4.4\%}$ \\
\cline{2-10} $\scriptstyle 4\,\mathrm{GeV} < p_{\mathrm{T}jet} < 6\,\mathrm{GeV}$
& \footnotesize{0.85 -- 0.9}  &  ${}_{-0.03\%}^{+1.6\%}$ & $\pm 1.2\%$ & ${}_{-6.5\%}^{+10\%}$ 
& $<0.005\%$ & $\pm 2.9\%$ & $\pm 0.02\%$ & $\pm 2.1\%$ & ${}_{-7.5\%}^{+11\%}$ \\
\hline \multirow{2}{*}{$\frac{1}{N_{jet}}\frac{dN_{ch}}{d\varptrel}$}
& \footnotesize{0.0 -- 0.05}  &  ${}_{-1.1\%}^{+1.2\%}$ & $\pm 0.10\%$ & ${}_{-3.0\%}^{+3.4\%}$ 
& ${}_{-0.17\%}^{+0.21\%}$ & $\pm 0.58\%$ & $\pm 0.05\%$ & $\pm 0.05\%$ & ${}_{-3.2\%}^{+3.6\%}$ \\
\cline{2-10}
& \footnotesize{0.5 -- 0.55}  &  $\pm 1.2\%$ & $\pm 0.01\%$ & ${}_{-2.4\%}^{+3.3\%}$ & ${}_{-0.29\%}^{+0.38\%}$ 
& $\pm 0.40\%$ & $\pm 0.03\%$ & $\pm 0.01\%$ & ${}_{-2.7\%}^{+3.5\%}$ \\
\cline{2-10} $\scriptstyle |y_{jet}| < 1.9$
& \footnotesize{0.85 -- 0.9}  &  ${}_{-1.0\%}^{+0.95\%}$ & $\pm 0.06\%$ & ${}_{-2.6\%}^{+3.3\%}$ 
& ${}_{-0.25\%}^{+0.32\%}$ & $\pm 0.57\%$ & $\pm 0.05\%$ & $\pm 0.09\%$ & ${}_{-2.9\%}^{+3.5\%}$ \\
\cline{2-10} $\scriptstyle 24\,\mathrm{GeV} < p_{\mathrm{T}jet} < 40\,\mathrm{GeV}$
& \footnotesize{3.0 -- 3.5}  &  ${}_{-2.4\%}^{+2.6\%}$ & $\pm 0.56\%$ & ${}_{-1.1\%}^{+8.7\%}$ 
& ${}_{-0.04\%}^{+0.05\%}$ & $\pm 0.27\%$ & $\pm 0.14\%$ & $\pm 0.48\%$ & ${}_{-2.7\%}^{+9.1\%}$ \\
\hline
\end{tabular}
\end{center}
\end{table*}

\begin{table*}[htb!]
\begin{center}

\caption{ Summary of systematic uncertainties for selected bins in selected multiplicity, $\varz$, and $\rho_{ch}(r)$ distributions, for $R$ = 0.6.
\label{table-systsummary-unitless}}
\vspace { 2mm }
\begin{tabular}{| c | c | c | c | c | c | c | c | c | c |}
\hline Distribution & Bin & Track & Track & Monte & High-$p_\mathrm{T}$ & Unmatched & Split & Clo- & Total   \\
 & & eff. & res. & Carlo & Tracks & Jets/Tracks & Vertex & sure &  \\
\hline \multirow{2}{*}{$\frac{1}{N_{jet}}\frac{dN_{jet}}{dN^{ch}_{jet}}$}
& \footnotesize{1}   &  ${}_{-6.3\%}^{+6.8\%}$ & $\pm 0.01\%$ & ${}_{-10\%}^{+7.9\%}$ & $<0.005\%$ 
& $\pm 0.15\%$ & $\pm 0.21\%$ & $\pm 0.25\%$ & ${}_{-12\%}^{+10\%}$ \\
\cline{2-10}
& \footnotesize{3}  &  ${}_{-2.6\%}^{+2.8\%}$ & $\pm 0.14\%$ & ${}_{-0.27\%}^{+3.3\%}$ & $<0.005\%$ 
& $\pm 0.16\%$ & $\pm 0.08\%$ & $\pm 0.26\%$ & ${}_{-2.6\%}^{+4.3\%}$ \\
\cline{2-10} $\scriptstyle |y_{jet}| < 1.9$
& \footnotesize{5}  &  ${}_{-1.3\%}^{+1.2\%}$ & $\pm 0.03\%$ & ${}_{-2.4\%}^{+0.23\%}$ & $<0.005\%$ 
& $\pm 0.03\%$ & $\pm 0.04\%$ & $\pm 0.09\%$ & ${}_{-2.7\%}^{+1.2\%}$ \\
\cline{2-10} $\scriptstyle 4\,\mathrm{GeV} < p_{\mathrm{T}jet} < 6\,\mathrm{GeV}$
& \footnotesize{9}  &  ${}_{-10\%}^{+11\%}$ & $\pm 0.36\%$ & ${}_{-8.4\%}^{+11\%}$ & $<0.005\%$ 
& $\pm 0.01\%$ & $\pm 0.24\%$ & $\pm 1.4\%$ & ${}_{-13\%}^{+15\%}$ \\
\hline \multirow{2}{*}{$\frac{1}{N_{jet}}\frac{dN_{jet}}{dN^{ch}_{jet}}$}
& \footnotesize{1}  &  ${}_{-11\%}^{+19\%}$ & $\pm 3.2\%$ & ${}_{-8.5\%}^{+50\%}$ & ${}_{-5.5\%}^{+3.7\%}$ 
& $\pm 0.13\%$ & $\pm 0.35\%$ & $\pm 35\%$ & ${}_{-38\%}^{+64\%}$ \\
\cline{2-10}
& \footnotesize{3}  &  ${}_{-9.9\%}^{+10\%}$ & $\pm 0.05\%$ & ${}_{-3.3\%}^{+3.1\%}$ & ${}_{-1.9\%}^{+1.7\%}$ 
& $\pm 0.14\%$ & $\pm 0.28\%$ & $\pm 0.77\%$ & $\pm 11\%$ \\
\cline{2-10} $\scriptstyle |y_{jet}| < 1.9$
& \footnotesize{5}  &  ${}_{-6.5\%}^{+6.8\%}$ & $\pm 0.03\%$ & ${}_{-1.4\%}^{+3.3\%}$ & ${}_{-0.75\%}^{+0.64\%}$ 
& $\pm 0.06\%$ & $\pm 0.21\%$ & $\pm 0.26\%$ & ${}_{-6.7\%}^{+7.6\%}$ \\
\cline{2-10} $\scriptstyle 24\,\mathrm{GeV} < p_{\mathrm{T}jet} < 40\,\mathrm{GeV}$
& \footnotesize{9}  &  ${}_{-1.7\%}^{+1.6\%}$ & $\pm 0.14\%$ & ${}_{-0.21\%}^{+2.9\%}$ & $\pm 0.07\%$ 
& $\pm 0.02\%$ & $\pm 0.08\%$ & $<0.005\%$ & ${}_{-1.7\%}^{+3.3\%}$ \\
\hline \multirow{2}{*}{$\frac{1}{N_{jet}}\frac{dN_{ch}}{d\varz}$}
& \footnotesize{0.1 -- 0.125}  &  $\pm 1.7\%$ & $\pm 0.11\%$ & ${}_{-8.9\%}^{+1.5\%}$ & $<0.005\%$ 
& $\pm 2.0\%$ & $\pm 0.05\%$ & $\pm 0.03\%$ & ${}_{-9.3\%}^{+3.0\%}$ \\
\cline{2-10}
& \footnotesize{0.5 -- 0.525}  &  ${}_{-1.3\%}^{+1.2\%}$ & $\pm 0.07\%$ & ${}_{-0.54\%}^{+3.2\%}$ & $<0.005\%$ 
& $\pm 2.0\%$ & $\pm 0.03\%$ & $\pm 0.18\%$ & ${}_{-2.4\%}^{+3.9\%}$ \\
\cline{2-10} $\scriptstyle |y_{jet}| < 1.9$
& \footnotesize{0.85 -- 0.9}  &  ${}_{-4.7\%}^{+4.5\%}$ & $\pm 0.20\%$ & ${}_{-4.0\%}^{+3.7\%}$ 
& $<0.005\%$ & $\pm 0.25\%$ & $\pm 0.12\%$ & $\pm 0.33\%$ & ${}_{-6.1\%}^{+5.9\%}$ \\
\cline{2-10} $\scriptstyle 4\,\mathrm{GeV} < p_{\mathrm{T}jet} < 6\,\mathrm{GeV}$
& \footnotesize{1.0}  &  ${}_{-6.5\%}^{+7.0\%}$ & $\pm 0.08\%$ & ${}_{-16\%}^{+4.8\%}$ & $<0.005\%$ 
& $\pm 0.03\%$ & $\pm 0.23\%$ & $\pm 0.28\%$ & ${}_{-17\%}^{+8.5\%}$ \\
\hline \multirow{2}{*}{$\frac{1}{N_{jet}}\frac{dN_{ch}}{d\varz}$}
& \footnotesize{0.1 -- 0.125}  &  ${}_{-1.3\%}^{+1.2\%}$ & $\pm 0.14\%$ & ${}_{-5.8\%}^{+3.0\%}$ & ${}_{-0.32\%}^{+0.42\%}$ 
& $\pm 0.01\%$ & $\pm 0.03\%$ & $\pm 0.09\%$ & ${}_{-5.9\%}^{+3.3\%}$ \\
\cline{2-10}
& \footnotesize{0.5 -- 0.525}  &  ${}_{-3.6\%}^{+3.9\%}$ & $\pm 0.29\%$ & ${}_{-0.86\%}^{+2.7\%}$ & ${}_{-0.65\%}^{+0.40\%}$ 
& $\pm 0.64\%$ & $\pm 0.07\%$ & $\pm 0.24\%$ & ${}_{-3.9\%}^{+4.8\%}$ \\
\cline{2-10} $\scriptstyle |y_{jet}| < 1.9$
& \footnotesize{0.85 -- 0.9}  &  ${}_{-9.3\%}^{+9.0\%}$ & $\pm 1.7\%$ & ${}_{-6.6\%}^{+1.8\%}$ & ${}_{-3.7\%}^{+3.3\%}$ 
& $\pm 0.21\%$ & $\pm 0.20\%$ & $\pm 1.3\%$ & ${}_{-12\%}^{+10\%}$ \\
\cline{2-10} $\scriptstyle 24\,\mathrm{GeV} < p_{\mathrm{T}jet} < 40\,\mathrm{GeV}$
& \footnotesize{0.95 -- 1.0}  &  $\pm 13\%$ & $\pm 1.3\%$ & ${}_{-3.5\%}^{+6.8\%}$ & ${}_{-5.5\%}^{+3.8\%}$ 
& $\pm 0.81\%$ & $\pm 0.30\%$ & $\pm 3.1\%$ & ${}_{-15\%}^{+16\%}$ \\
\hline \multirow{2}{*}{$\rho_{ch}(r)$}
& \footnotesize{0.0 -- 0.01}  &  ${}_{-6.8\%}^{+7.1\%}$ & $\pm 0.12\%$ & ${}_{-46\%}^{+20\%}$ & $<0.005\%$ 
& $\pm 0.30\%$ & $\pm 0.22\%$ & $\pm 0.36\%$ & ${}_{-46\%}^{+21\%}$ \\
\cline{2-10}
& \footnotesize{0.09 -- 0.1}  &  ${}_{-0.06\%}^{+0.00\%}$ & $\pm 0.22\%$ & ${}_{-10\%}^{+2.6\%}$ & $<0.005\%$ 
& $\pm 2.4\%$ & $<0.005\%$ & $\pm 0.12\%$ & ${}_{-10\%}^{+3.5\%}$ \\
\cline{2-10} $\scriptstyle |y_{jet}| < 1.9$
& \footnotesize{0.28 -- 0.3}  &  ${}_{-1.1\%}^{+0.90\%}$ & $\pm 0.11\%$ & ${}_{-10\%}^{+2.5\%}$ & $<0.005\%$ 
& $\pm 2.4\%$ & $\pm 0.03\%$ & $\pm 0.03\%$ & ${}_{-11\%}^{+3.6\%}$ \\
\cline{2-10} $\scriptstyle 4\,\mathrm{GeV} < p_{\mathrm{T}jet} < 6\,\mathrm{GeV}$
& \footnotesize{0.55 -- 0.6}  &  ${}_{-2.8\%}^{+2.4\%}$ & $\pm 0.06\%$ & ${}_{-3.1\%}^{+0.94\%}$ & $<0.005\%$ 
& $\pm 3.8\%$ & $\pm 0.07\%$ & $\pm 5.4\%$ & ${}_{-7.8\%}^{+7.1\%}$ \\
\hline \multirow{2}{*}{$\rho_{ch}(r)$}
& \footnotesize{0.0 -- 0.01} &  ${}_{-3.4\%}^{+3.3\%}$ & $\pm 0.08\%$ & ${}_{-5.1\%}^{+4.2\%}$ & ${}_{-0.59\%}^{+0.48\%}$ 
& $\pm 0.20\%$ & $\pm 0.09\%$ & $\pm 0.20\%$ & ${}_{-6.2\%}^{+5.4\%}$ \\
\cline{2-10}
& \footnotesize{0.09 -- 0.1} &  ${}_{-0.60\%}^{+0.56\%}$ & $\pm 0.04\%$ & ${}_{-4.6\%}^{+2.9\%}$ & ${}_{-0.26\%}^{+0.33\%}$ 
& $\pm 0.21\%$ & $<0.005\%$ & $\pm 0.05\%$ & ${}_{-4.7\%}^{+3.0\%}$ \\
\cline{2-10} $\scriptstyle |y_{jet}| < 1.9$
& \footnotesize{0.28 -- 0.3}  &  ${}_{-2.1\%}^{+2.0\%}$ & $\pm 0.07\%$ & ${}_{-8.0\%}^{+3.4\%}$ & ${}_{-0.35\%}^{+0.45\%}$ 
& $\pm 0.86\%$ & $\pm 0.10\%$ & $\pm 0.04\%$ & ${}_{-8.4\%}^{+4.1\%}$ \\
\cline{2-10} $\scriptstyle 24\,\mathrm{GeV} < p_{\mathrm{T}jet} < 40\,\mathrm{GeV}$
& \footnotesize{0.55 -- 0.6} &  ${}_{-2.3\%}^{+2.5\%}$ & $\pm 0.02\%$ & ${}_{-6.7\%}^{+3.8\%}$ & ${}_{-0.36\%}^{+0.47\%}$ 
& $\pm 1.3\%$ & $\pm 0.15\%$ & $\pm 0.64\%$ & ${}_{-7.2\%}^{+4.8\%}$ \\
\hline
\end{tabular}
\end{center}
\end{table*}

\begin{description}
\item[Tracking efficiency and resolution]
The systematic uncertainties associated with tracking efficiency and resolution 
are computed by applying efficiency scaling and resolution smearing to single tracks in generated events.  
The efficiencies and resolutions are derived from fully-simulated events
and parameterized by a track's $1/p_\mathrm{T}$, $\eta$, and the $\Delta R$ to the nearest other track.  
These efficiency uncertainties are determined by adjusting the derived single-track efficiencies 
in accordance with uncertainties determined from data~\cite{:2010ir}, for which uncertainties in the material description of the ATLAS inner detector dominate.
The resulting distributions are compared with those found for the baseline resolution, with the increase (decrease) taken as the upper (lower) systematic uncertainty. 
Similarly, the uncertainties in tracking resolution are investigated by applying an additional Gaussian resolution smearing
derived from studies of high-momentum muons; the resulting change is taken as a symmetric uncertainty.

\item[Monte Carlo]
The efficiency scaling and resolution smearing are similarly used for uncertainties associated with the particular event generator used to compute the unfolding corrections.
For each tune described in Section~\ref{mc}, an ensemble of 50 samples with the same number of events as the data has the smearing applied,
and the average unfolded distribution is compared to one produced similarly (with 196 data-sized samples) for the central AMBT1 MC sample.
The largest increase (decrease) is taken as the upper (lower) systematic uncertainty.
Although the response matrix primarily models the impact of detector efficiency and resolution, truth-level details can impact the corrections in several ways.
Different jet or track momentum distributions change the population \emph{within} bins;
differences in the amount or distribution of activity in the underlying event can significantly change the track momentum and radius
 distribution, especially in the low-momentum bins,
because these tracks are not correlated with the jet direction and so may appear at large radii.
Differing strangeness fractions can change the distribution of long-lifetime tracks that decay and produce kinks in tracks,
leading to momentum mismeasurements and/or loss of tracks due to failed hit requirements.
The variation over tunes also accounts for PDF uncertainties, because MC properties are tuned with a particular PDF and several different PDF's are used.

\item[High-$p_\mathrm{T}$ Tracks]
Despite the tightening of the $d_0$ cut for high-momentum tracks,
0.6$\%$ (8.3$\%$) of selected tracks in the simulation with momentum above 10 $(40)\,\GeV$ do not have a matching truth particle with momentum within 50$\%$.
These high-momentum mismeasured tracks are primarily associated with wide-angle scatterings in the material of the inner detector that 
create the appearance of a single, straight track~\cite{:2010ir}.
Were the fraction of such tracks similar in data, their impact on the measurement would be accounted for in
the unfolding procedure. However, the data have a larger fraction of high-$p_\mathrm{T}$ tracks failing the $d_0$ cut (2.3$\%$) than does the MC (1.6$\%$).
A systematic uncertainty on the number of mismeasured tracks in data, as a function of track $p_\mathrm{T}$, is computed and propagated to assess its impact on jets.
The upper bound of this uncertainty assumes that all such
extra tracks are well-measured, but did not pass the $d_0$ cut due to improperly modeled resolution.
The lower bound is based on the assumption that the increase in rejected tracks corresponds to a proportional increase
in accepted mismeasured tracks.
These uncertainties are then propagated to each measured jet bin, with the scale factor determined 
by the fraction of jets in a given jet bin with their leading track in each momentum range.
Thus the correction is largest for high jet momentum
and low number of particles per jet, 
because these jets have the highest leading track momenta.
This uncertainty on the reconstructed data is then used to scale the measured distributions, with the unfolding applied in each case and compared to the unfolded central value.
The resulting
differences in each measured bin give the final uncertainty due to high-momentum mismeasured tracks.
This correction is also applied to the $\varz$, $p_{\mathrm{T}}^{rel}$ and $r$ distributions.  
Each bin is scaled in proportion to the fraction of tracks in that bin
that come from a jet whose leading track is in a given range, in proportion to the uncertainty on tracks in that range.  
The corrections are only significant for high-$\varz$ bins in high-$p_\mathrm{T}$ jets.
The corrections are then propagated through the unfolding as in the jet-based distributions, and the result compared to the central value.

\item[Unmatched jets and tracks]
In order to assign an uncertainty to the correction for unmatched reconstructed jets and tracks, 
the unfolding is repeated with a correction determined from a fully-simulated Perugia 2010
in place of the baseline ATLAS AMBT1.
The difference in the unfolding output between this and the baseline configuration is taken to be the uncertainty on the correction.
The uncertainty is symmetrized by taking the maximum deviation.

\item[Split vertex] 
The data are corrected for event rejection due to misreconstruction of extra primary vertices 
as a function of the number of selected tracks in the event, as discussed in Section~\ref{seleff}.
A weight is applied on an event-by-event basis based on the probability of that event being rejected. 
As the correction is derived entirely from simulation, the full value of the correction is taken as a (symmetrized) uncertainty.

\item[Closure]
As discussed in Section~\ref{analysis}, closure tests reveal a bias in the correction procedure.  This is taken
as a bin-by-bin systematic uncertainty, which is symmetrized.
\end{description}

The total uncertainty on the cross section is dominated by tracking efficiency, with mismeasured high-momentum tracks also playing a role for the highest-momentum jets.
Tracking efficiency uncertainties play a similar role for the multiplicity distributions, especially for low-multiplicity jets, with
Monte Carlo uncertainties also making large contributions.
For other distributions, Monte Carlo uncertainties are dominant, with closure uncertainties making large contributions in extremal bins;
tracking efficiency uncertainties mostly cancel because these distributions are normalized by the number of jets.

\section{Results and Discussion}

\begin{figure*}[htb!]
\begin{center}
\subfigure[$\ R$ = 0.4 Distribution]{\label{fig:paper_ntrk4}\includegraphics[width=0.48\textwidth]{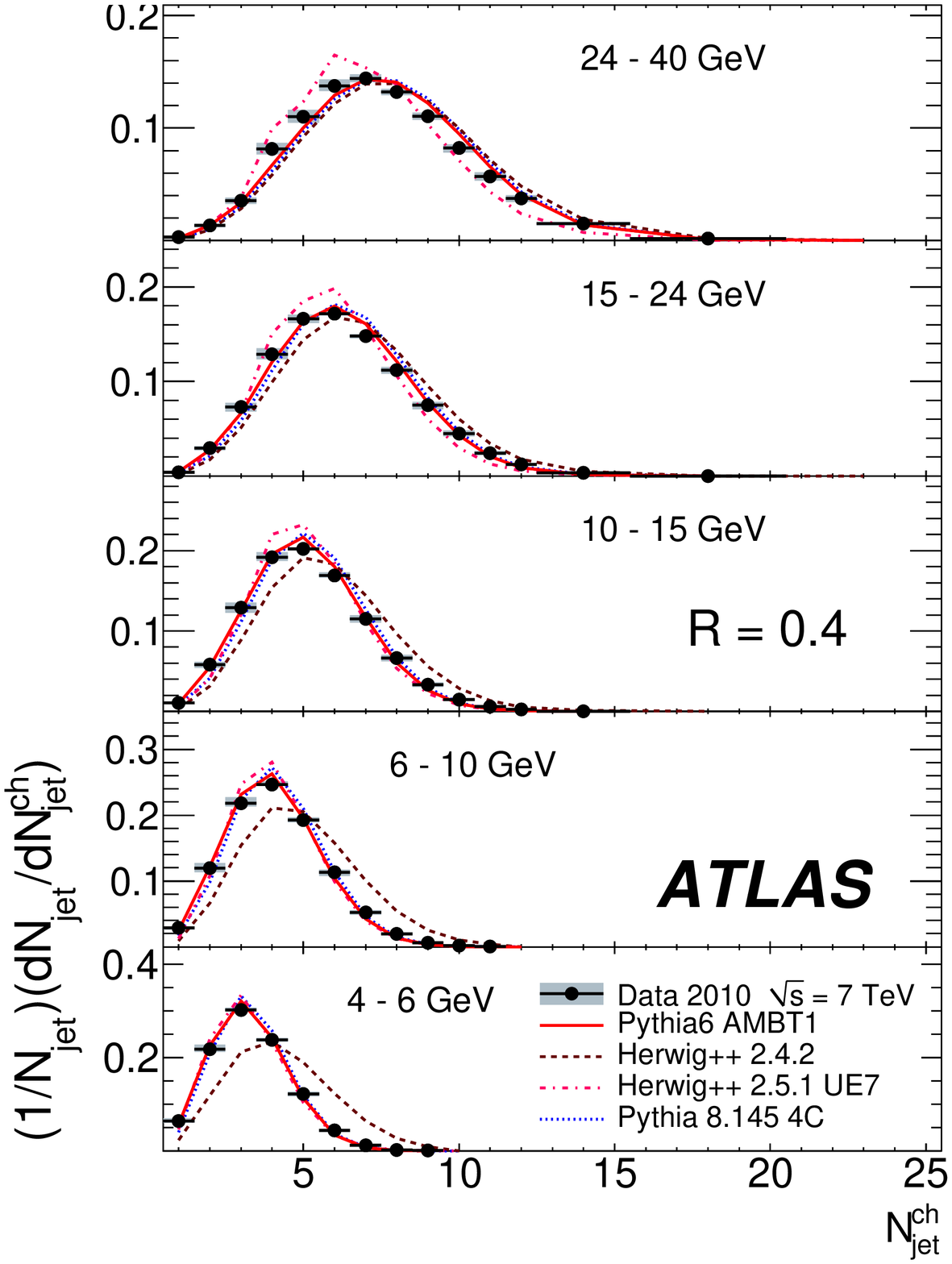}}
\subfigure[$\ R$ = 0.4 Data-MC Difference]{\label{fig:paper_ntrk4_rat}\includegraphics[width=0.48\textwidth]{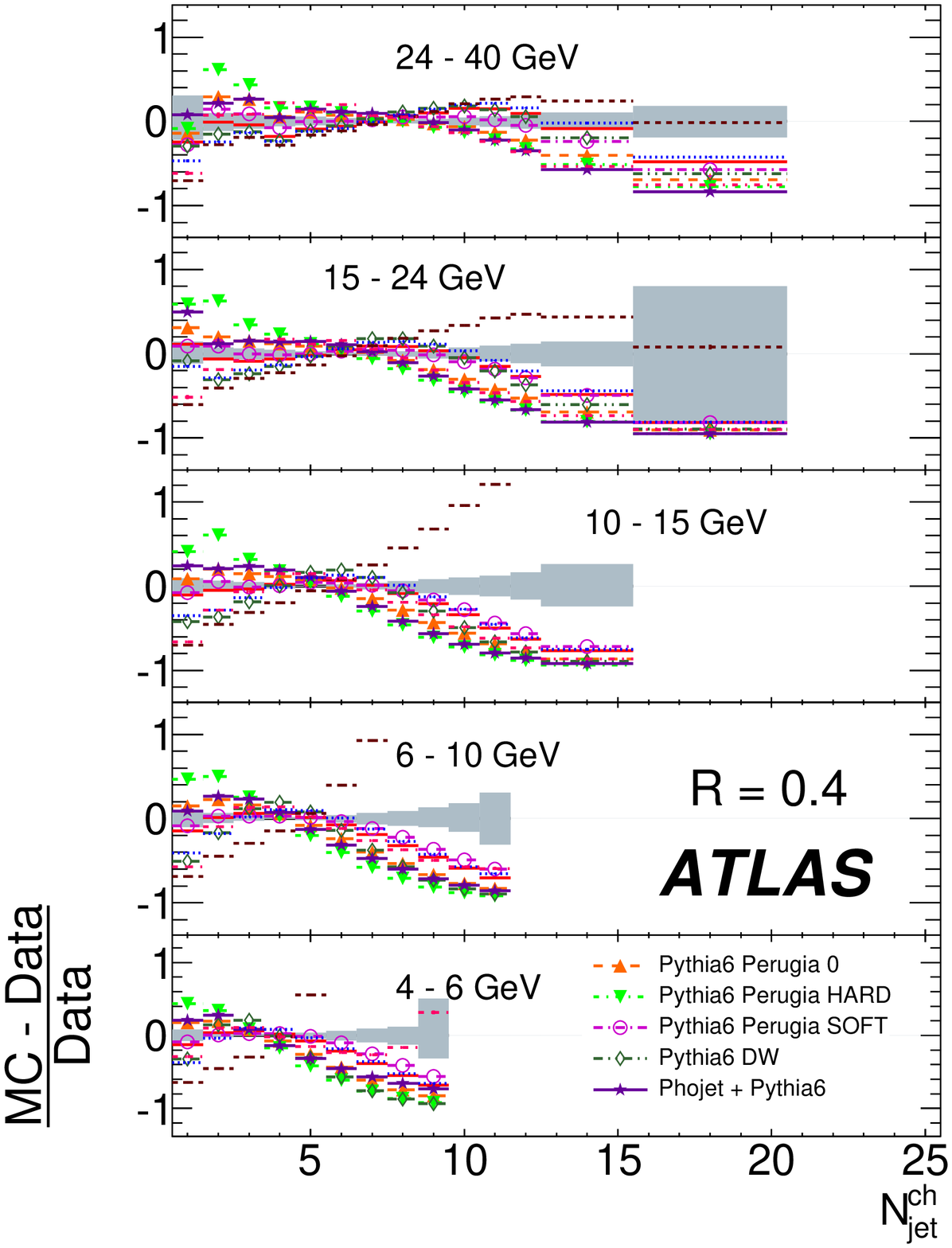}}
\subfigure[$\ R$ = 0.6 Distribution]{\label{fig:paper_ntrk6}\includegraphics[width=0.48\textwidth]{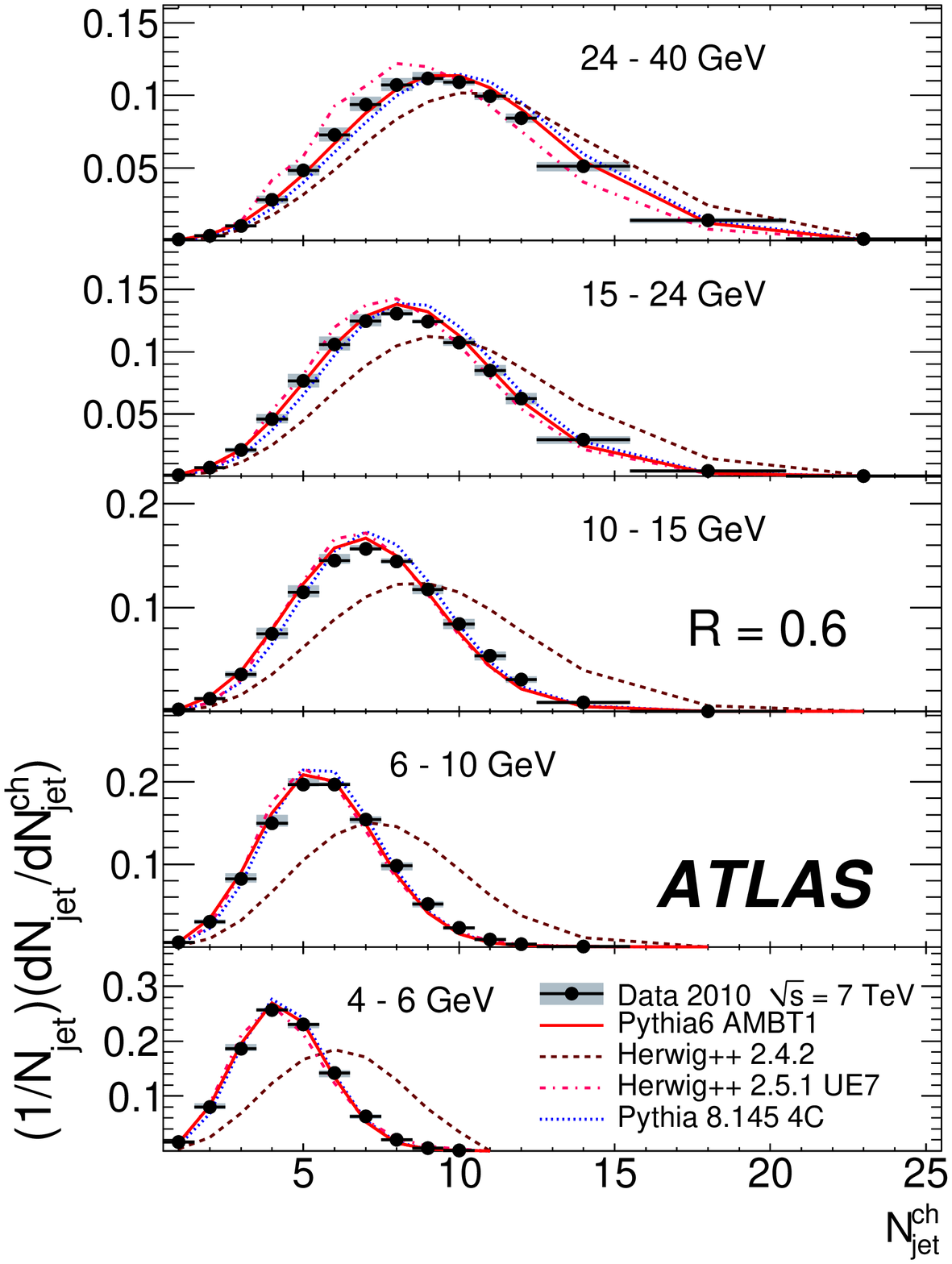}}
\subfigure[$\ R$ = 0.6 Data-MC Difference]{\label{fig:paper_ntrk6_rat}\includegraphics[width=0.48\textwidth]{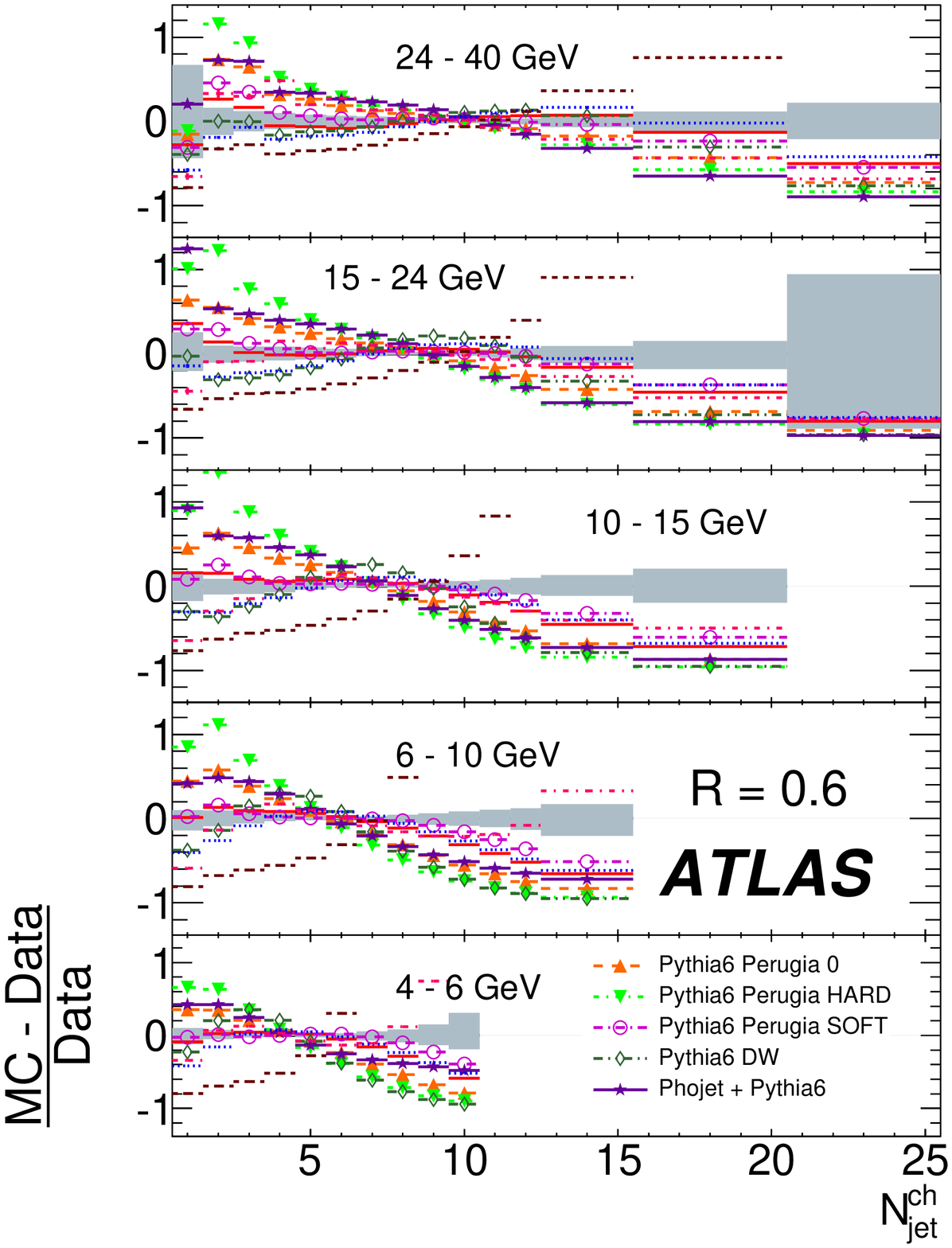}}
\caption{
Multiplicity of particles per charged particle jet, over the full measured rapidity range $|y| < 1.9$, with anti-$\kt$ radius parameter $R$ as indicated.
Figure a (c) shows the distributions for five momentum ranges with $R$ = 0.4 (0.6), and Figure b (d) shows the fractional difference
between a range of Monte Carlo event generator predictions and the data.
The distributions for the Perugia 2010 tune, not shown, agree qualitatively with the Perugia 0 tune.
\label{fig:paper_ntrk}}
\end{center}
\end{figure*}

\begin{figure*}[htb!]
\begin{center}
\subfigure[$\ R$ = 0.4 Distribution]{\label{fig:paper_z4_cs}\includegraphics[width=0.48\textwidth]{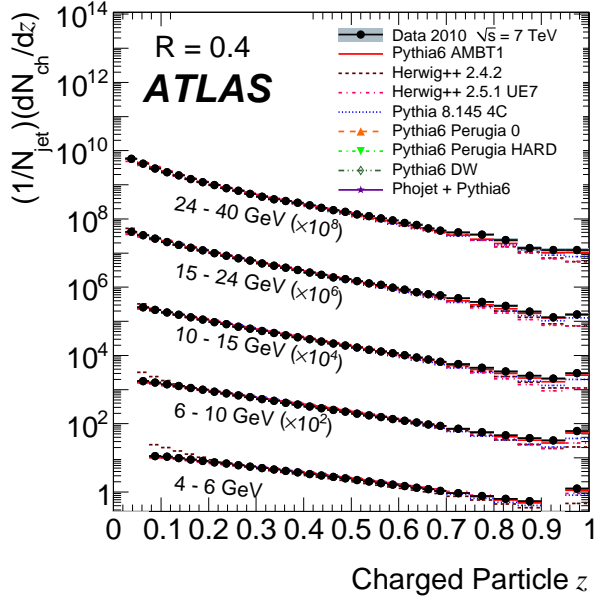}}
\subfigure[$\ R$ = 0.4 Data-MC Difference]{\label{fig:paper_z4_rat}\includegraphics[width=0.48\textwidth]{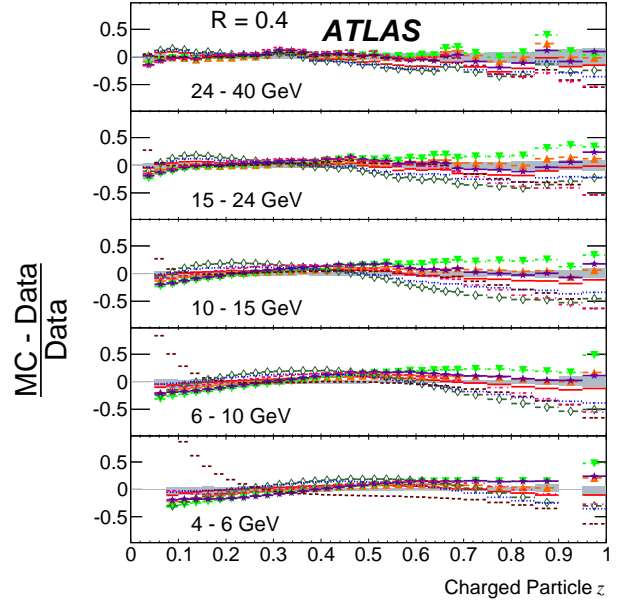}}
\subfigure[$\ R$ = 0.6 Distribution]{\label{fig:paper_z6_cs}\includegraphics[width=0.48\textwidth]{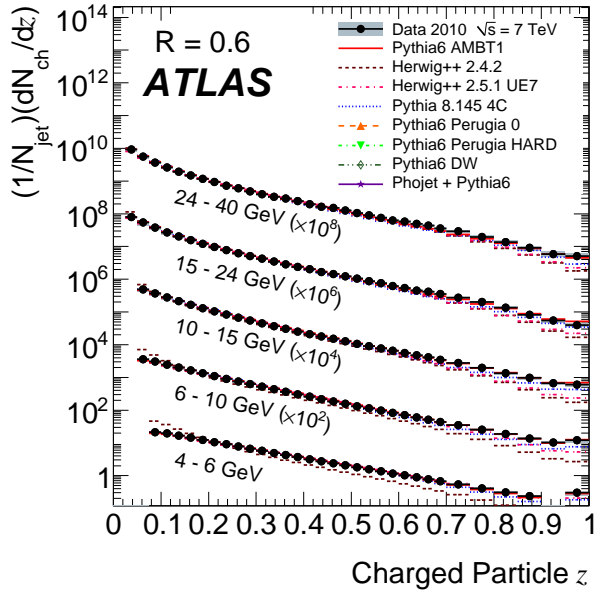}}
\subfigure[$\ R$ = 0.6 Data-MC Difference]{\label{fig:paper_z6_rat}\includegraphics[width=0.48\textwidth]{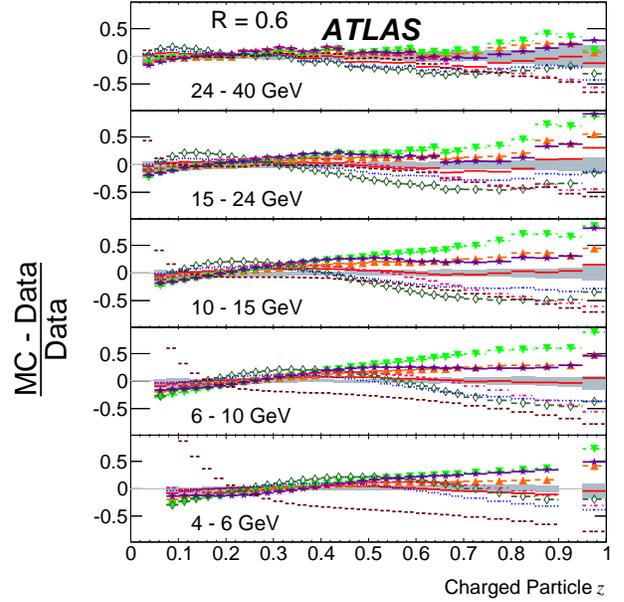}}
\caption{
The distribution of the fragmentation variable $z$ for anti-$\kt$ jets with radius parameter $R$ as indicated, in the rapidity range $|y| < 1.9$.
Figure a (c) shows the distributions for five momentum ranges with $R$ = 0.4 (0.6), and Figure b (d) shows the fractional difference
between a range of Monte Carlo event generator predictions and the data.
The distributions for the Perugia 2010 (Perugia SOFT) tune, not shown, agree qualitatively with the Perugia 0 (AMBT1) tune.
\label{fig:paper_z}}
\end{center}
\end{figure*}

\begin{figure*}[htb!]
\begin{center}
\subfigure[$\ R$ = 0.4 Distribution]{\label{fig:paper_ptrel4_cs}\includegraphics[width=0.48\textwidth]{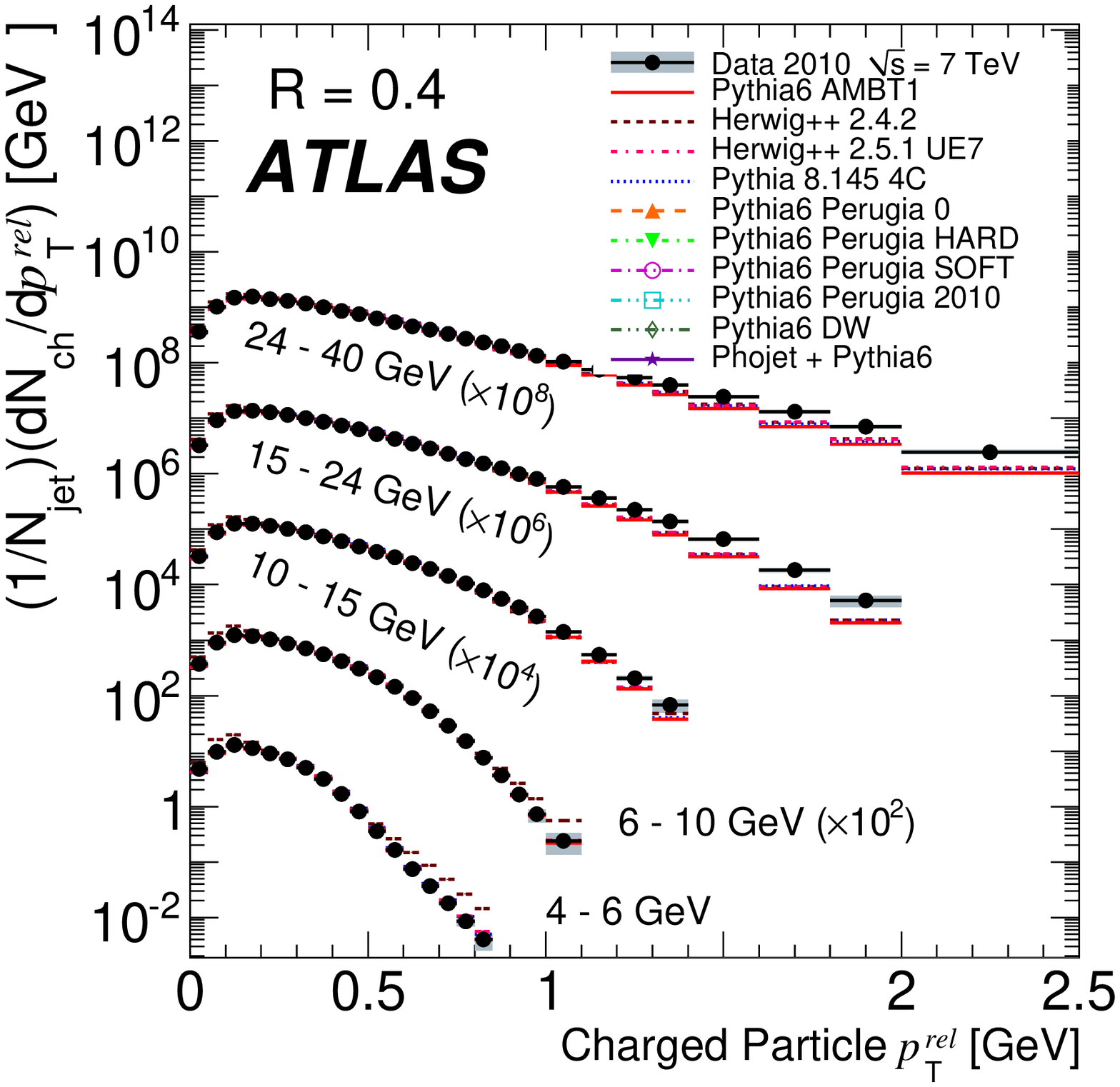}}
\subfigure[$\ R$ = 0.4 Data-MC Difference]{\label{fig:paper_ptrel4_rat}\includegraphics[width=0.48\textwidth]{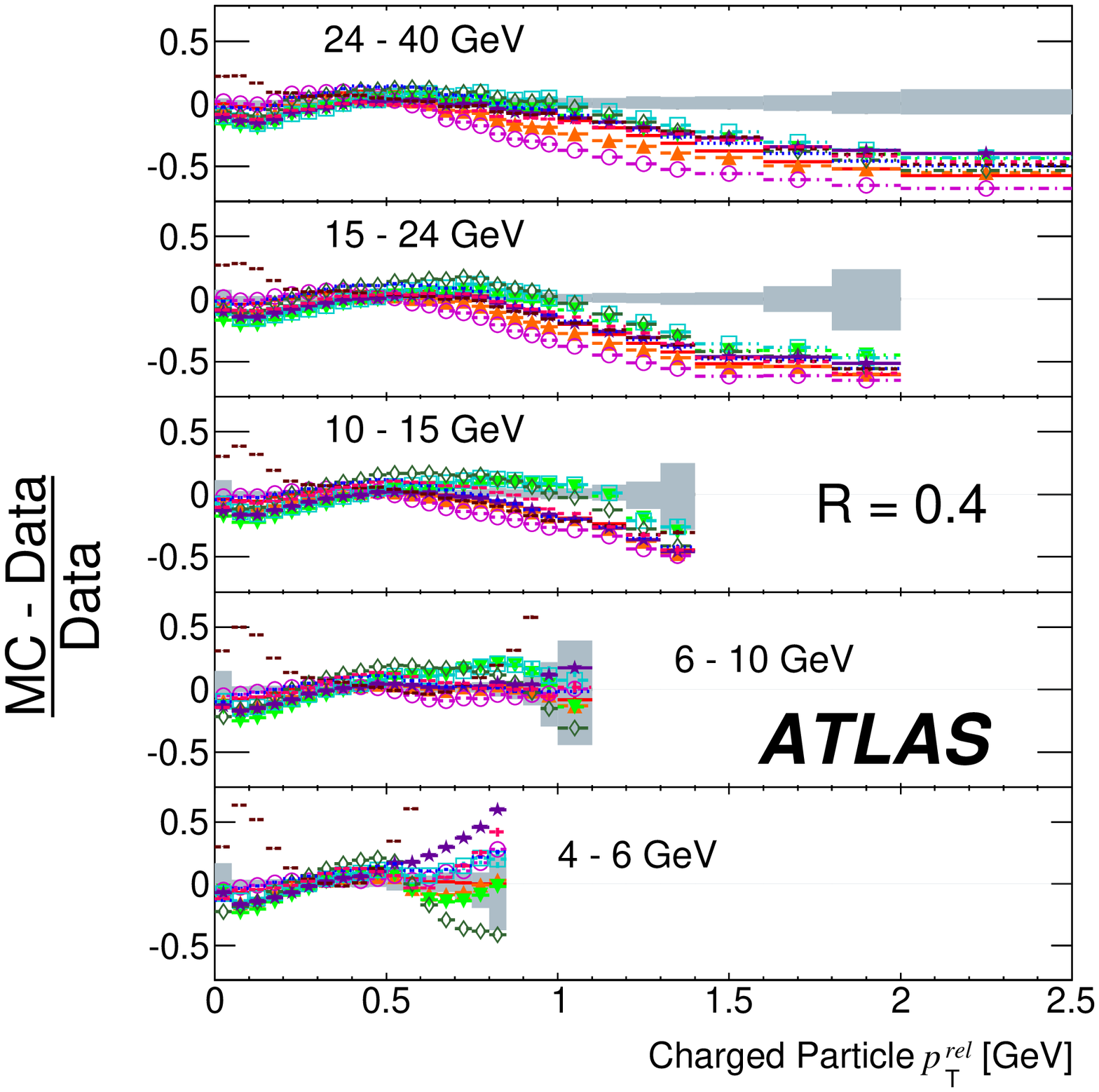}}
\subfigure[$\ R$ = 0.6 Distribution]{\label{fig:paper_ptrel6_cs}\includegraphics[width=0.48\textwidth]{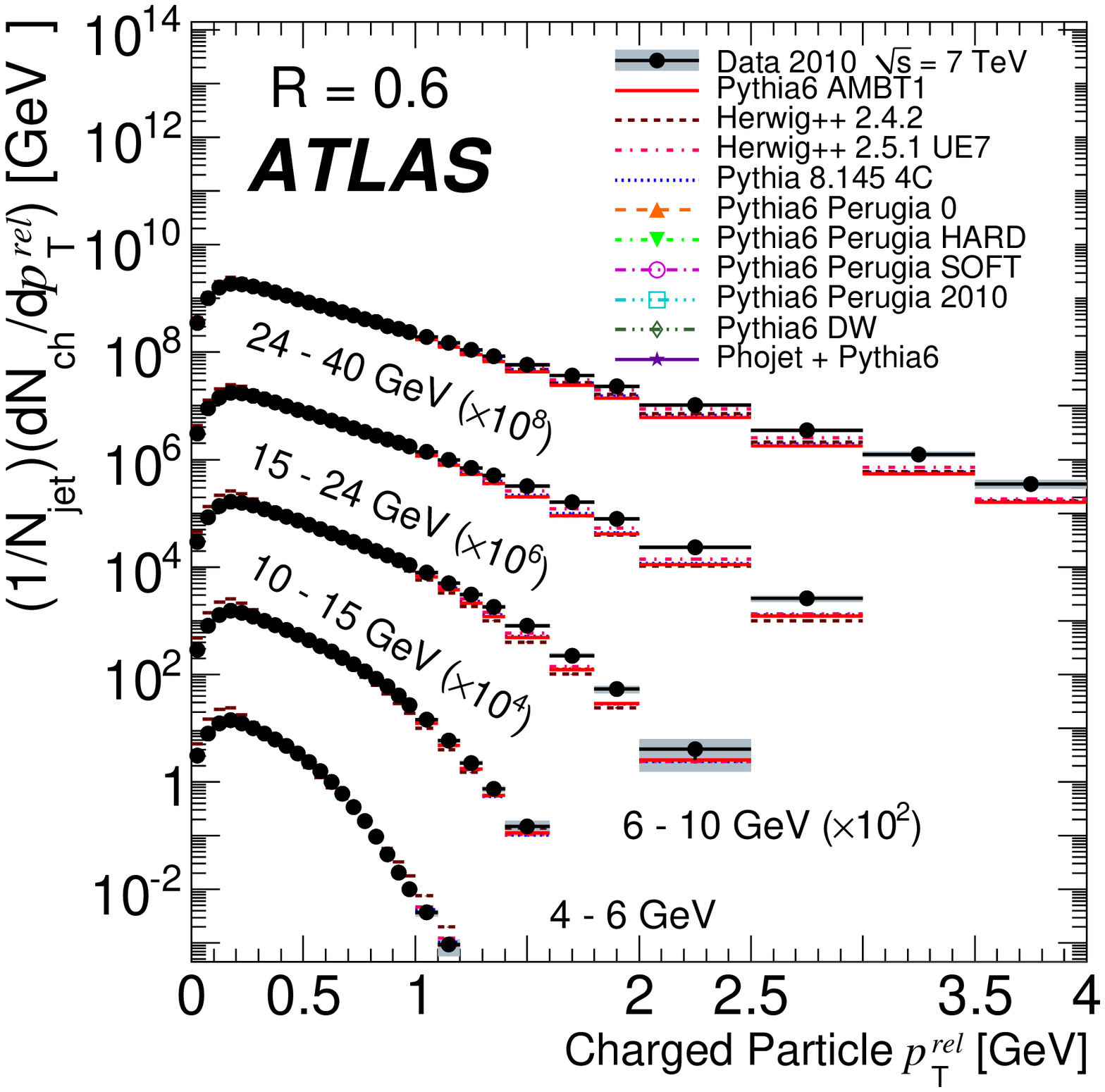}}
\subfigure[$\ R$ = 0.6 Data-MC Difference]{\label{fig:paper_ptrel6_rat}\includegraphics[width=0.48\textwidth]{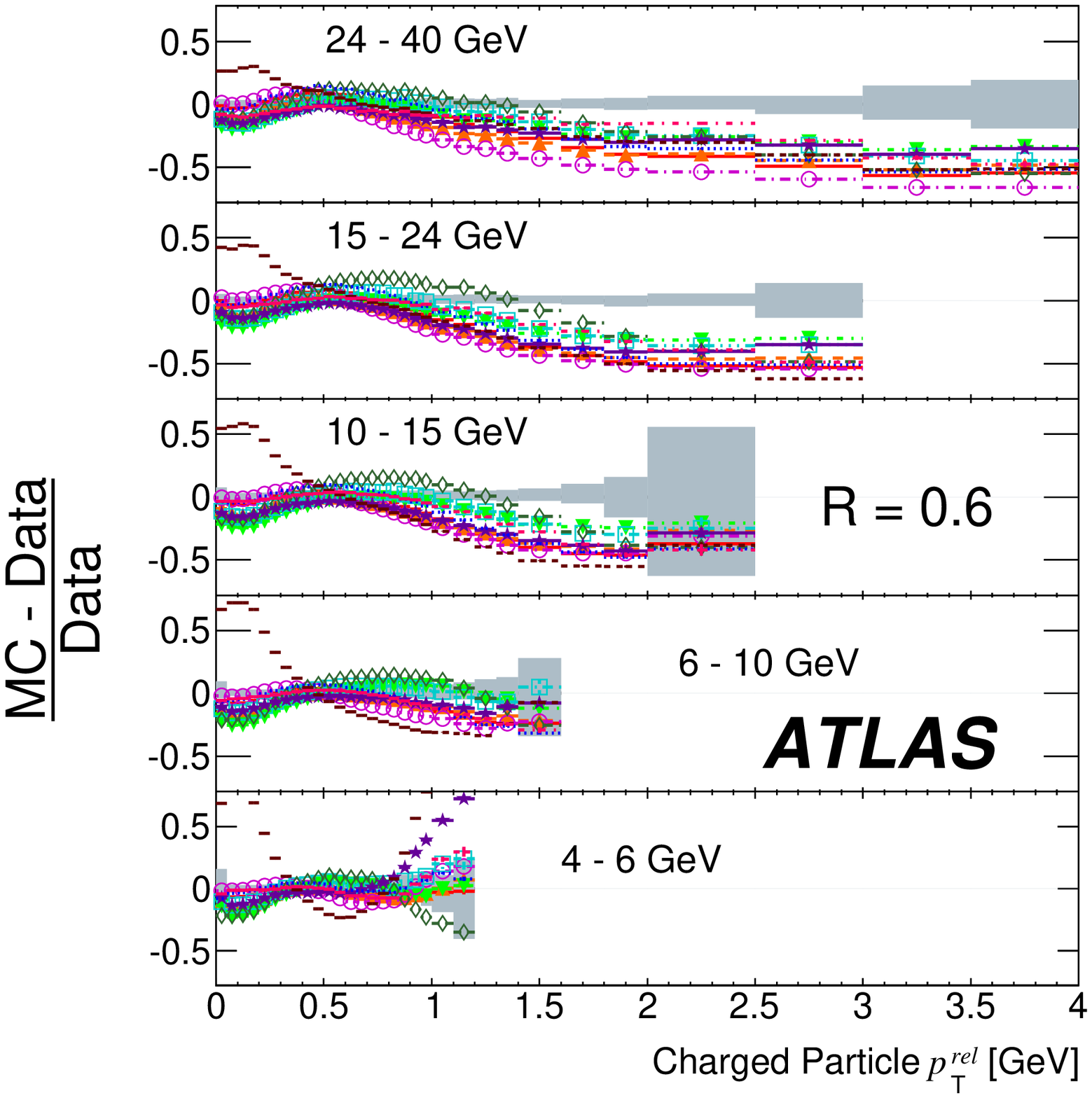}}
\caption{
The distribution of the charged particle transverse momentum $p_{\mathrm{T}}^{rel}$ with respect to anti-$\kt$ jets with radius parameter $R$ as indicated, 
in the rapidity range $|y| < 1.9$.
Figure a (c) shows the distributions for five momentum ranges with $R$ = 0.4 (0.6), and Figure b (d) shows the fractional difference
between a range of Monte Carlo event generator predictions and the data.
\label{fig:paper_ptrel}}
\end{center}
\end{figure*}

\begin{figure*}[htb!]
\begin{center}
\subfigure[$\ R$ = 0.4 Distribution]{\label{fig:paper_rdA4_cs}\includegraphics[width=0.48\textwidth]{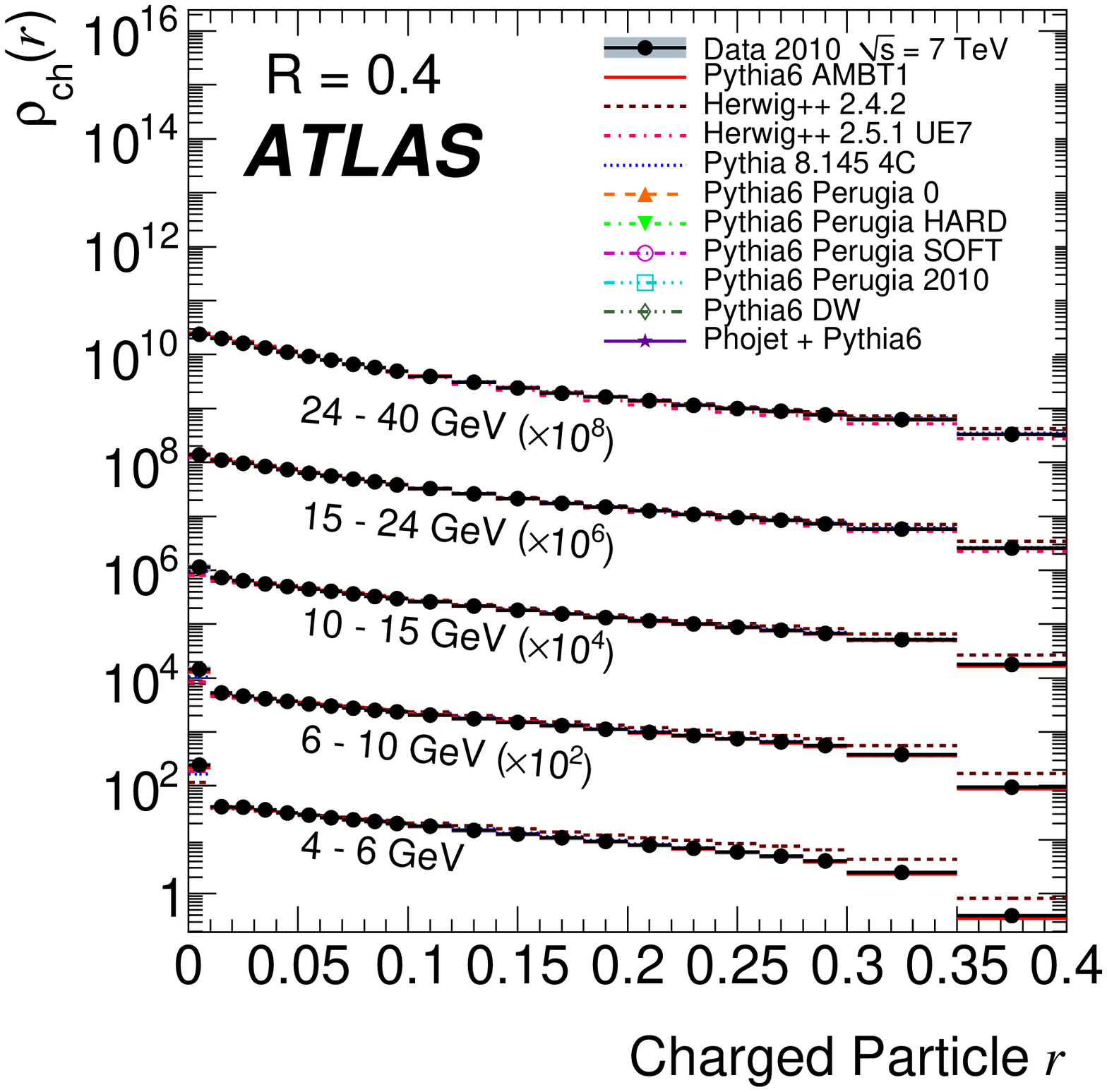}}
\subfigure[$\ R$ = 0.4 Data-MC Difference]{\label{fig:paper_rdA4_rat}\includegraphics[width=0.48\textwidth]{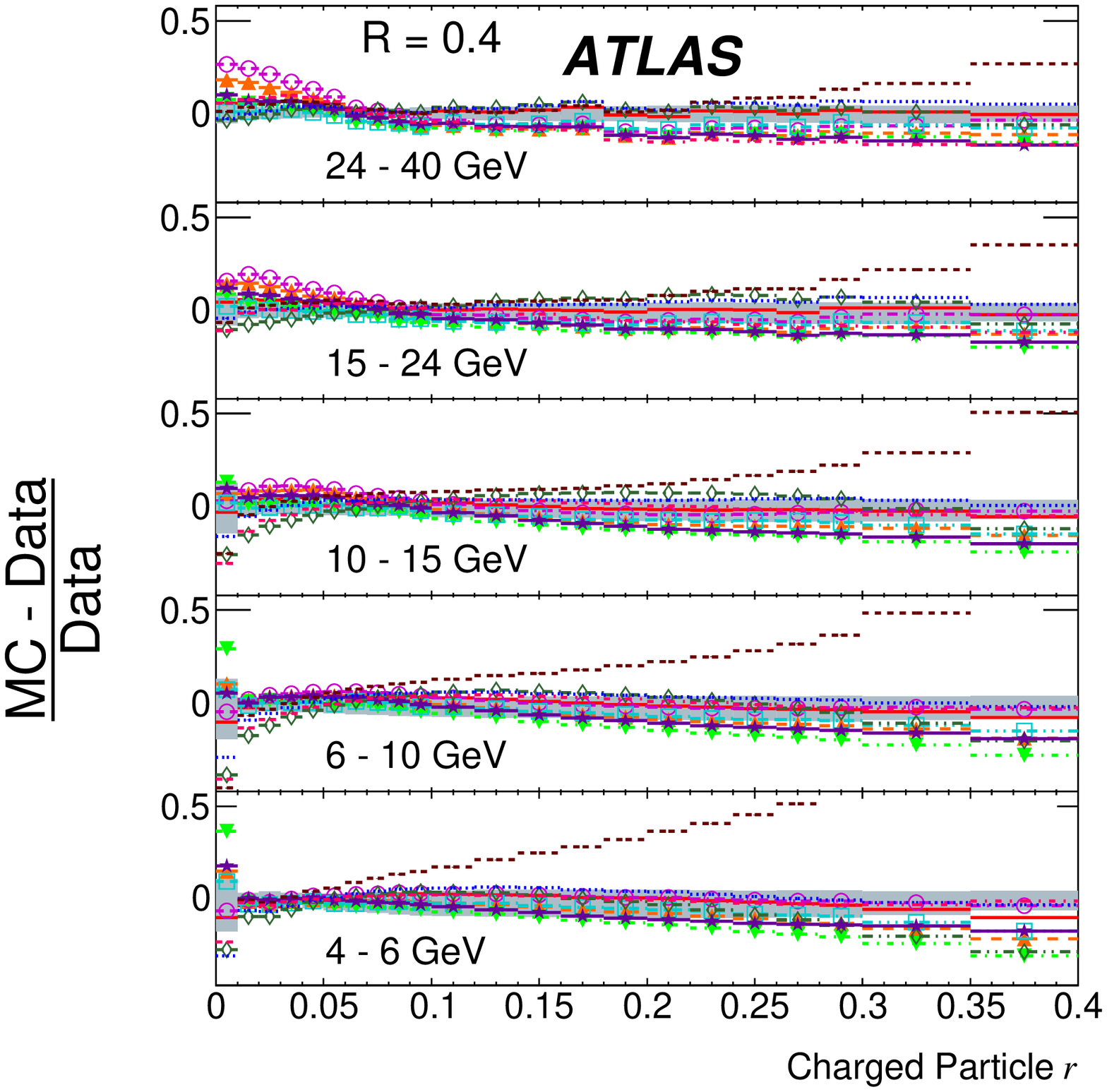}}
\subfigure[$\ R$ = 0.6 Distribution]{\label{fig:paper_rdA6_cs}\includegraphics[width=0.48\textwidth]{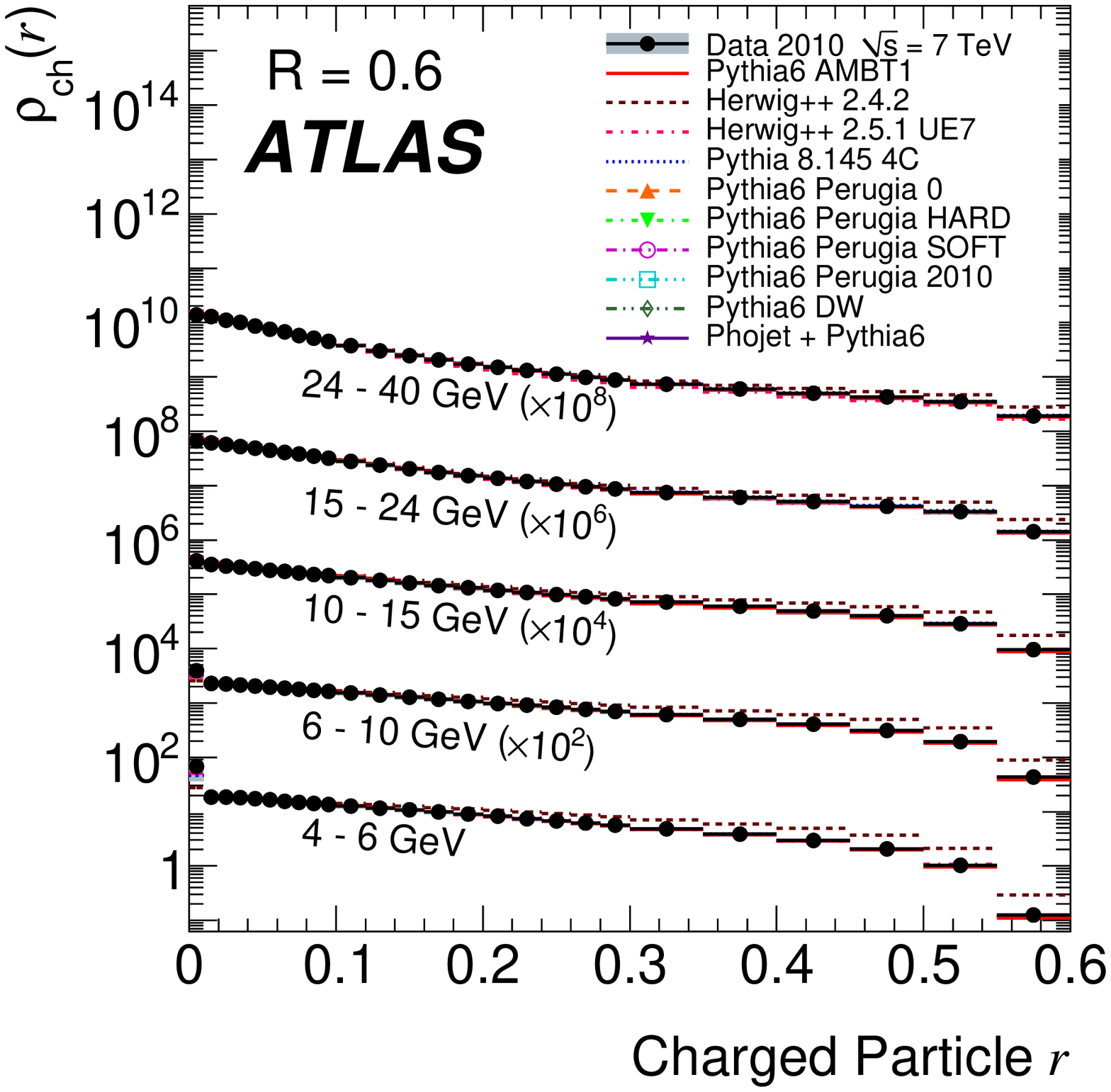}}
\subfigure[$\ R$ = 0.6 Data-MC Difference]{\label{fig:paper_rdA6_rat}\includegraphics[width=0.48\textwidth]{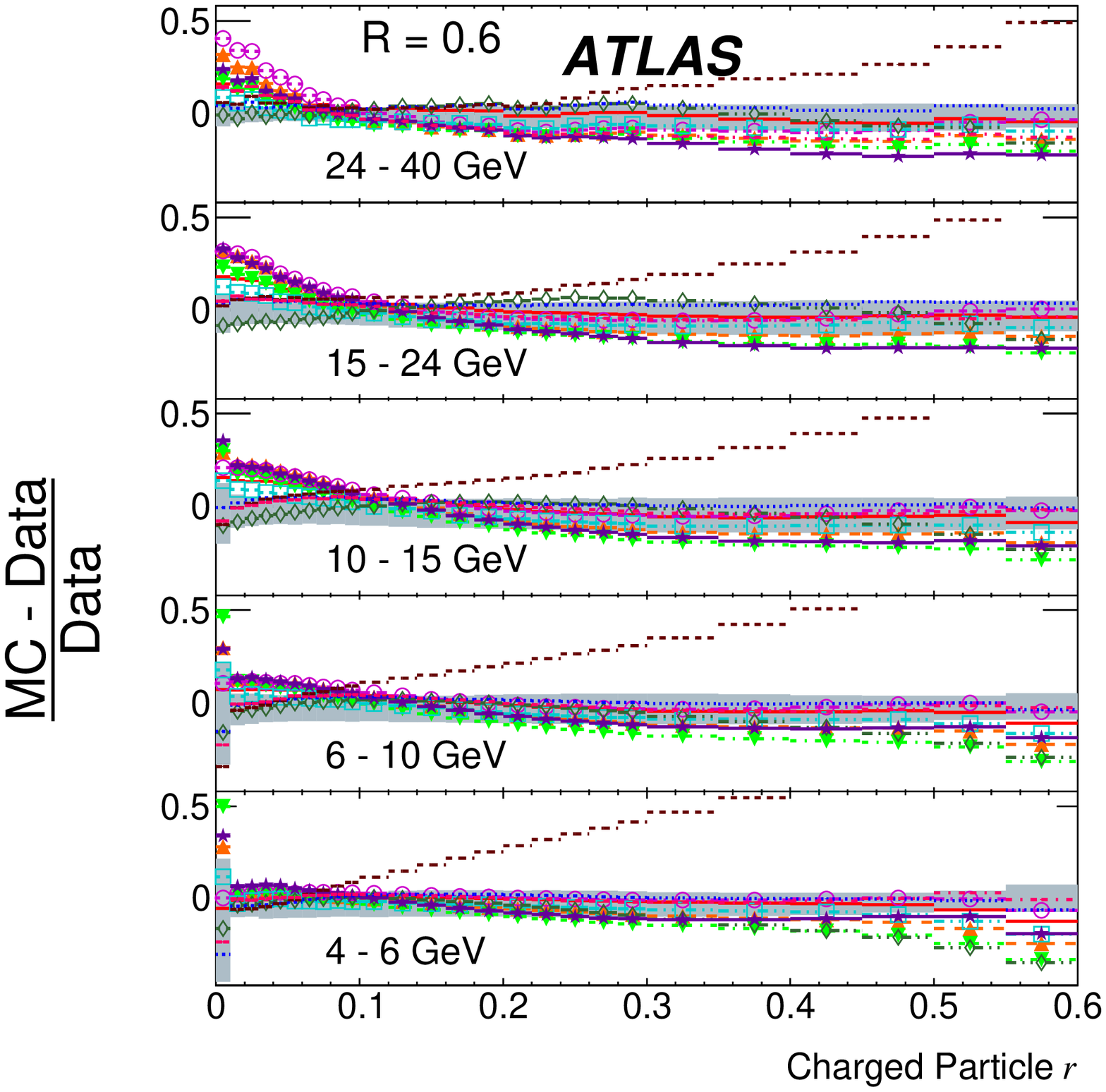}}
\caption{
The distribution of the charged particle number density $\rho_{ch}(r)$ for anti-$\kt$ jets with radius parameter $R$ as indicated, in the rapidity range $|y| < 1.9$.
Figure a (c) shows the distributions for five momentum ranges with $R$ = 0.4 (0.6), and Figure b (d) shows the fractional difference
between a range of Monte Carlo event generator predictions and the data.
\label{fig:paper_rdA}}
\end{center}
\end{figure*}

A selection of the distributions measured in this analysis appears in Figures~\ref{fig:paper_pt}--\ref{fig:paper_rdA}.  
Other rapidity ranges may be found in Ref.~\footnote{ 
A complete set of tables for all distributions in all measured rapidity bins are available
at the Durham HepData repository (\url{http://hepdata.cedar.ac.uk}).
}.
They are compared to the MC distributions described in Section~\ref{mc}. 

\subsection{Charged particle jet cross section}

\begin{figure*}[htb!]
\begin{center}
\subfigure[$\ R$ = 0.4]{\label{fig:paper_pt4}\includegraphics[width=0.48\textwidth]{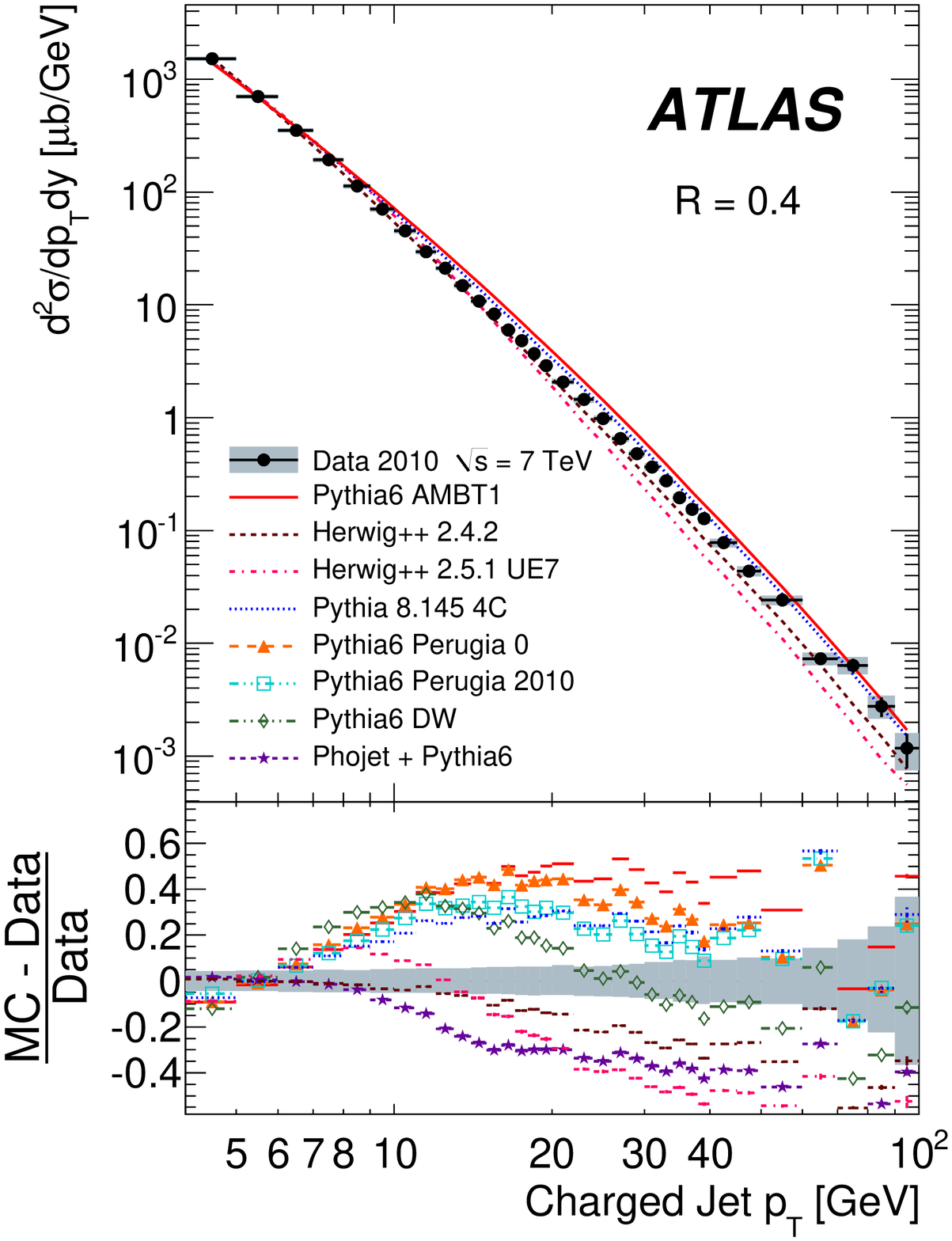}}
\subfigure[$\ R$ = 0.6]{\label{fig:paper_pt6}\includegraphics[width=0.48\textwidth]{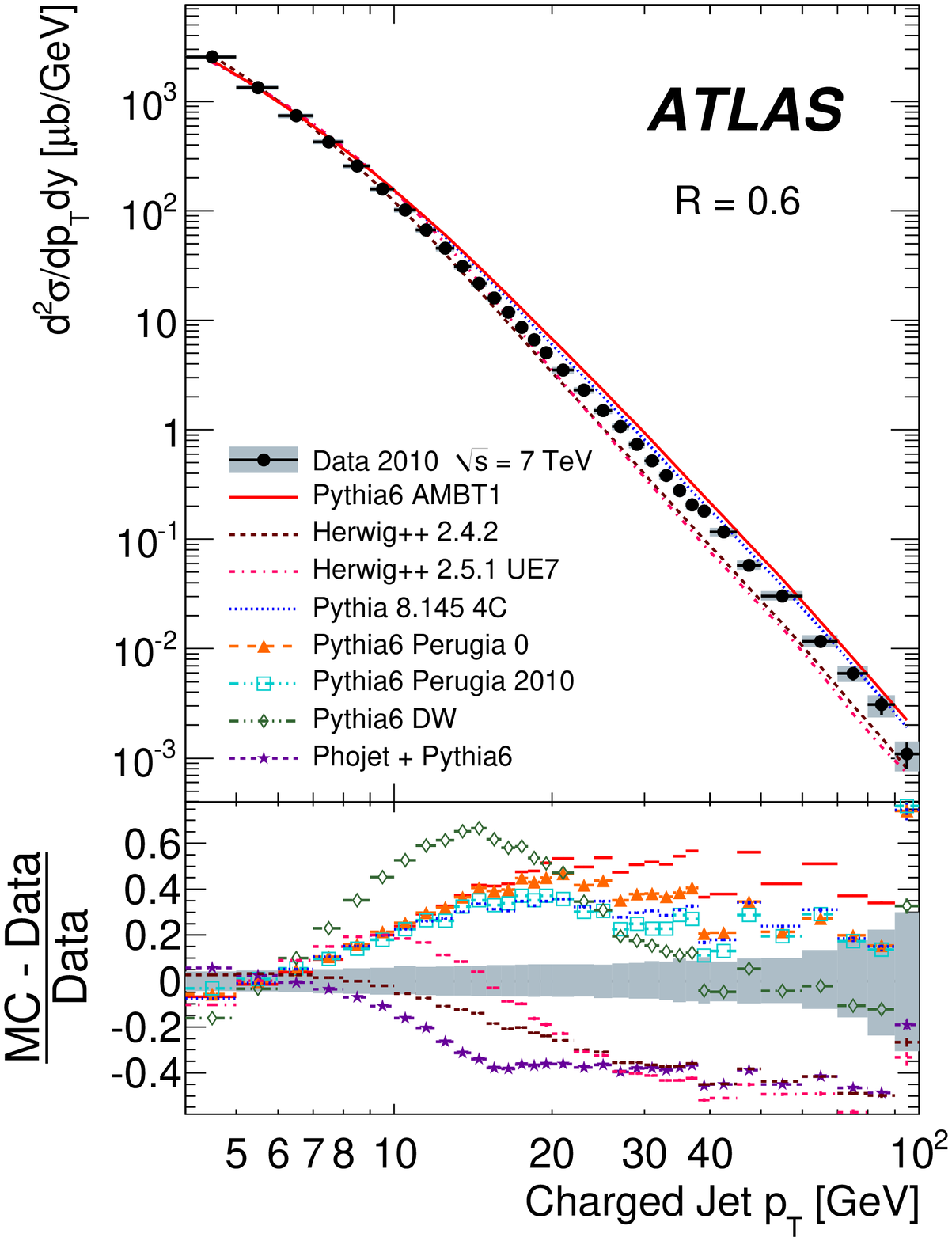}}
\caption{
The cross section for anti-$\kt$ charged particle jets as a function of $p_\mathrm{T}$, with $|y| < 0.5$ and radius parameter $R$ as indicated.
The shaded area is the total uncertainty for the corrected data distribution, excluding the overall 3.4$\%$ luminosity uncertainty.
The data are compared to a range of theoretical results from Monte Carlo event generators, which are normalized to the data over the full momentum and rapidity range measured,
using the scale factor $S$ as defined in the text.
The bottom inserts show the fractional difference between these distributions and the data.
The distributions for the Perugia HARD (Perugia SOFT) tune, not shown, agree qualitatively with the Perugia2010 (Perugia 0) tune.
\label{fig:paper_pt}}
\end{center}
\end{figure*}

Cross sections as a function of jet $p_\mathrm{T}$ are shown in Figure~\ref{fig:paper_pt}. 
The simulated cross sections shown for comparison are scaled to the data, using the scale factor $S$ defined by:

\begin{equation} S = \sigma^{total}_{data} / \sigma^{total}_{MC} \label{eqn:S} \end{equation}

\noindent where 

\begin{equation} \sigma^{total} = \int_{4\: \GeV}^{100\: \GeV} dp_{\mathrm{T}jet} \int_{-1.9}^{1.9} dy_{jet} \frac{d^2\sigma_{jet}}{dp_{\mathrm{T}jet}dy_{jet}} \end{equation}

\noindent The scale factors for the various MC's are given in Table~\ref{table-S}.

\begin{table}[htb!]
\caption{ Scale factors $S$ for Monte Carlo cross section normalization as defined in Equation~\ref{eqn:S}.  
The total Monte Carlo cross sections are normalized to the total for the data, over the full momentum and rapidity ranges
investigated in this analysis, $4\: \GeV < p_{\mathrm{T}jet} < 100\: \GeV$ and $|y_{jet}| < 1.9$. 
The statistical uncertainties on these ratios are less than $0.1\%$.
The systematic uncertainties are ${}^{+4.9\%}_{-4.8\%}$ ($\pm 5.0\%$) for $R$ = 0.4 (0.6);
these uncertainties are entirely correlated within columns, and largely correlated between columns.
\label{table-S}}
\vspace { 2mm }
\begin{tabular}{|c|c|c|}
\hline Tune & $S$ ($R$ = 0.4) & $S$ ($R$ = 0.6) \\
\hline AMBT1 & 0.838 & 0.896 \\
\hline Perugia-0 & 0.981 & 1.087 \\
\hline Perugia HARD & 0.936 & 1.058 \\
\hline Perugia SOFT & 0.968 & 1.036 \\
\hline Perugia 2010 & 0.976 & 1.044 \\
\hline DW & 0.894 & 1.045 \\
\hline Herwig++ 2.4.2 & 0.753 & 0.612 \\
\hline Herwig++ 2.5.1 UE7 & 0.425 & 0.458 \\ 
\hline Pythia 8 & 0.777 & 0.815 \\
\hline Phojet & 0.643 & 0.668 \\
\hline
\end{tabular}
\end{table}

\begin{figure*}[htb!]
\begin{center}
\subfigure[$\ R$ = 0.4]{\label{fig:paper_y4}\includegraphics[width=0.48\textwidth]{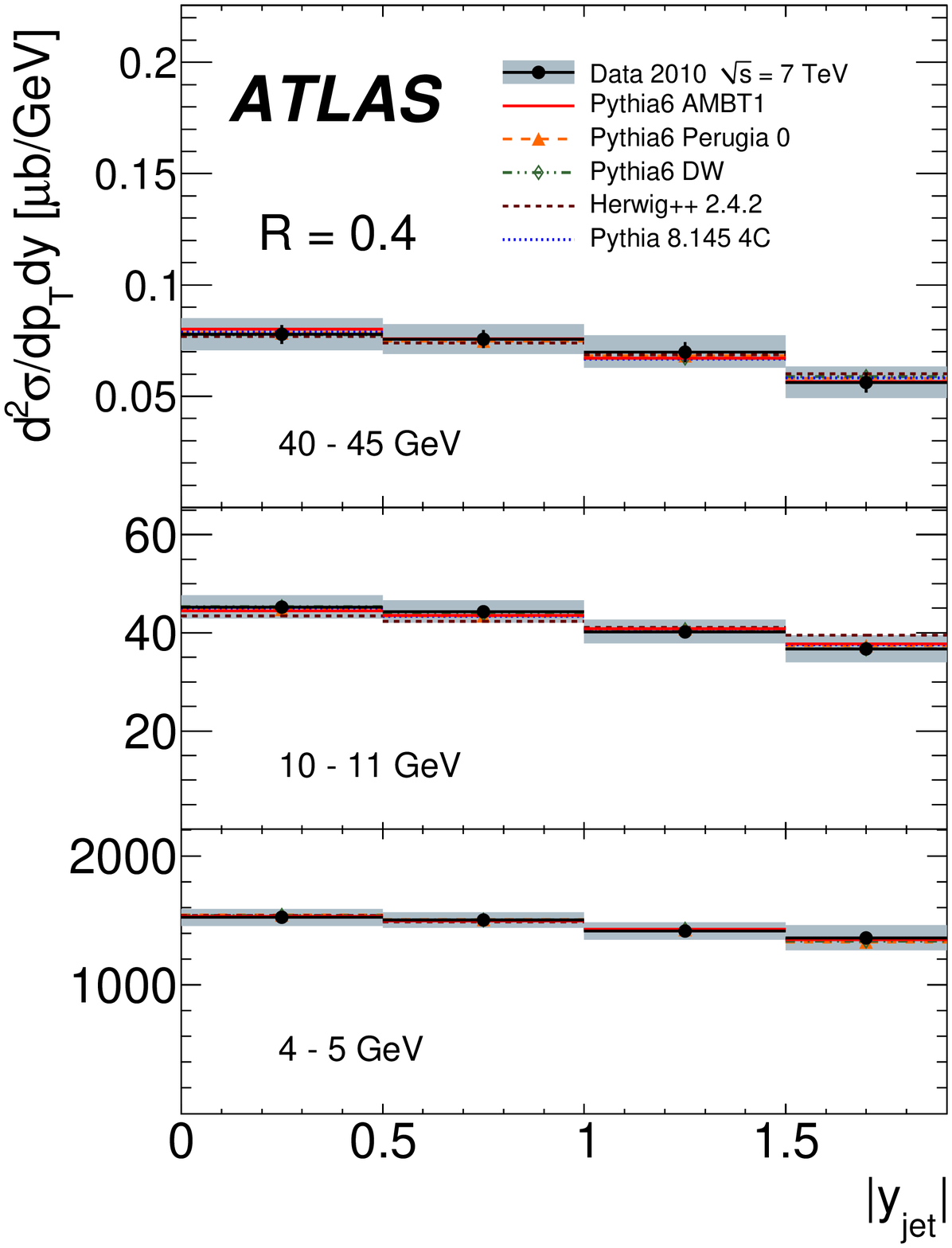}}
\subfigure[$\ R$ = 0.6]{\label{fig:paper_y6}\includegraphics[width=0.48\textwidth]{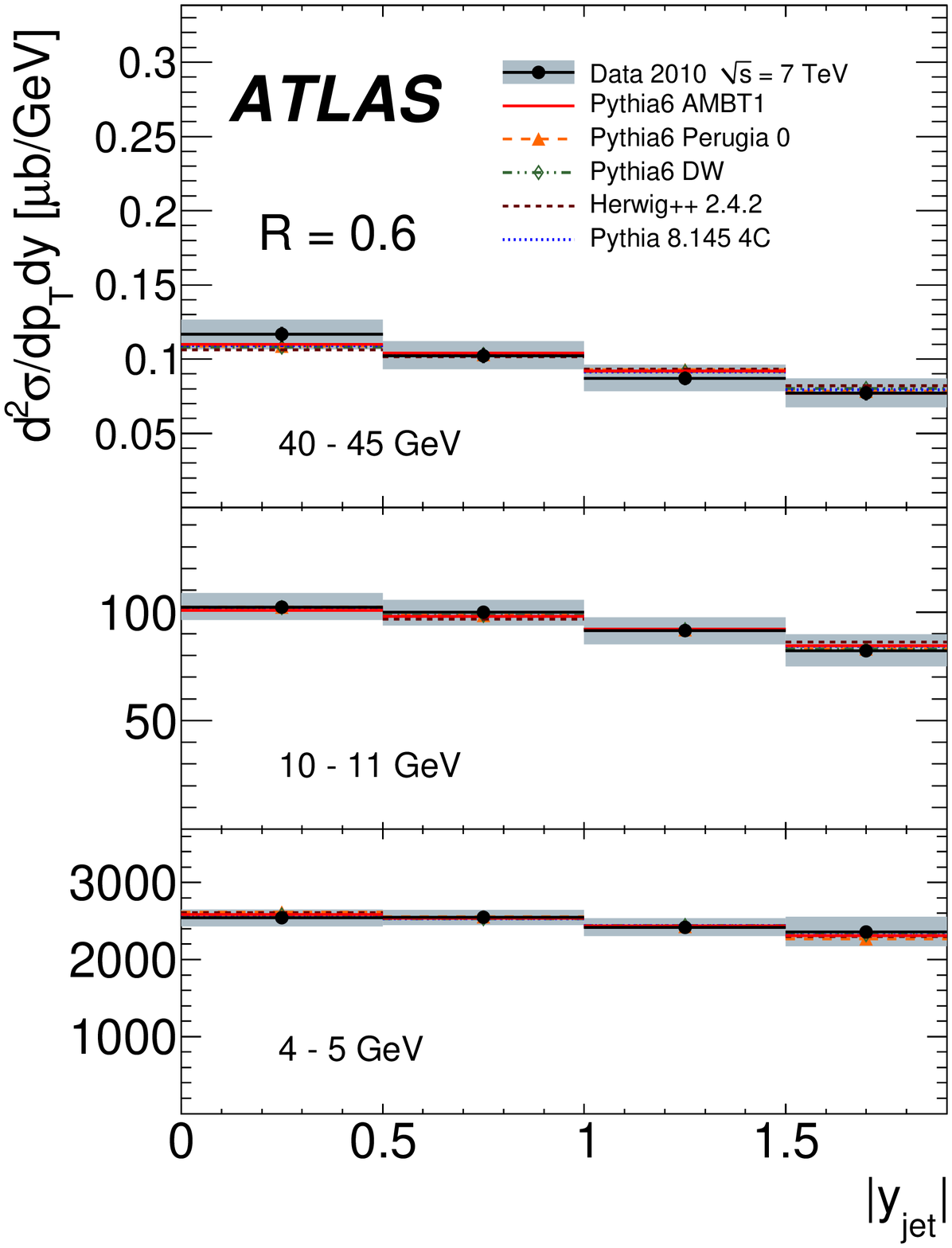}}
\caption{
The cross section for anti-$\kt$ charged particle jets as a function of rapidity, for selected momentum bins and radius parameter $R$ as indicated.
The shaded area is the total uncertainty for the corrected data distribution, excluding the overall 3.4$\%$ luminosity uncertainty.
The data are compared to a range of theoretical results from Monte Carlo event generators, which are normalized to the data separately for each momentum range.
The distributions for all Perugia samples agree qualitatively, so only Perugia 0 is shown.  The two Herwig++ tunes agree, so only Herwig++ 2.4.2 is shown.
\label{fig:paper_y}}
\end{center}
\end{figure*}

In all cases except Herwig++ 2.4.2, the scale factors $S$ for $R$ = 0.6 are larger than those for $R$ = 0.4.
Since the larger radius parameter results in the inclusion of more particles not directly associated with perturbative scattering,
this implies that the models underestimate the contribution of the underlying event required to reproduce the data.
Phojet has the best agreement between the scale factors at $R$ = 0.4 and $R$ = 0.6.
The Pythia Perugia tunes also agree well for the two radii, and are most consistent with one.

The jet cross section distributions (Fig.~\ref{fig:paper_pt}) fall by 6 orders of magnitude between jet momenta of 4 and $100\,\GeV$.  
The MC models considered agree broadly with this trend, but do not agree well in detail.
By construction of the normalization factor $S$, all distributions agree with the data in the lowest momentum bins;
most also give qualitative agreement for the shape at the lowest momenta.
The MC distributions diverge from the data in the 10--$20\,\GeV$ range, with some having a harder and others a softer momentum dependence,.
At higher $p_\mathrm{T}$, many of the models' momentum dependence agrees well with the data.
If one identifies the higher-momentum region as dominated by perturbative modeling and the low-momentum region as dominated by soft physics,
this indicates that perturbative modeling of charged particle jets is in fair agreement for most of the tunes.
It is the transition from soft physics to the perturbative region that is not successfully modeled.

The Pythia models give a harder shape for the momentum spectrum than the data below a jet $p_\mathrm{T}$ of about $20\,\GeV$, after which they exhibit roughly 
the same momentum dependence or become slightly softer.  
By contrast, the Phojet and Herwig++ models produce spectra that are softer than the data in the 10--$20\,\GeV$ range but have relatively good shape agreement outside,
although the Herwig++ 2.5.1 UE7 tune has the additional feature of producing too hard a spectrum at momenta below $10\,\GeV$.

The DW tune has a cross section that falls much too slowly in the transition region, 
and falls much too rapidly at high momentum; this effect is especially pronounced for $R$ = 0.6.  This suggests that the $Q^2$-ordered
 showering used by the DW tune is less successful in modeling the jet momentum spectrum.

Cross sections as a function of rapidity are shown in Figure~\ref{fig:paper_y}. 
The MC distributions are normalized to the data in each momentum bin separately rather than to the scale factor $S$.
The rapidity dependence of the cross section shows generally good agreement between data and MC.  
The cross section decreases only slightly with increasing rapidity at low momenta, but by a somewhat higher amount at higher momenta.

\subsection{Charged particle kinematics and multiplicity in jets}

The multiplicity of charged tracks per jet, for several momentum ranges over the full range $|y_{jet}| < 1.9$, is shown in Figure~\ref{fig:paper_ntrk}.
The charged particle $\varz$, $\varptrel$, and $\rho_{ch}(r)$ distributions are shown for same central rapidity range
in Figures~\ref{fig:paper_z}, \ref{fig:paper_ptrel}, and \ref{fig:paper_rdA} respectively.  

No tune describes well all of the  kinematic distributions and multiplicities of charged particles within jets. 
For $\varz$, the ATLAS AMBT1 and Perugia SOFT give good descriptions of the data.
For $\rho_{ch}(r)$, AMBT1 gives a good description.
For $p_\mathrm{T}^{rel}$ and multiplicity, no tune correctly describes all data.
For all distributions, Herwig++ 2.4.2 shows strong disagreement with the data, characterized by an excess of low-momentum particles, 
which is especially pronounced for the larger jet-finding parameter $R$ and at large particle $r$ (as defined in Equation~\ref{eqn:rdef}). 
Herwig++ 2.5.1 UE7 represents a significant improvement, 
so Herwig++ 2.4.2 will not be discussed further.

All remaining models give good agreement for the average charged particle multiplicity per jet.
The AMBT1 and Perugia SOFT tunes agree well with the multiplicity distributions (Figure~\ref{fig:paper_ntrk}) for the vast majority of jets, 
and Perugia 0, Perugia 2010, and Phojet give fair agreement for $R$ = 0.4,
although the high-multiplicity tail in data is greatly underestimated by all models.  
Herwig++ 2.5.1 UE7 and Perugia HARD have a significant excess of low-multiplicity jets, while Pythia 8.145 4C and Pythia DW exhibit a deficit.

The AMBT1 and Perugia SOFT tunes give good agreement with the measured longitudinal momentum fraction $\varz$ (Figure~\ref{fig:paper_z});
Perugia-0 also agrees well for $R$ = 0.4.
The other MC's (except Herwig++ 2.4.2) agree within $30\%$ at low $\varz$, but diverge more significantly at high $\varz$.
Perugia HARD has the most significant excess of high-$\varz$ particles, with excesses also present for Phojet and Perugia 0. 
The excess is particularly large at lower jet momenta and $R$ = 0.6,
suggesting that the soft physics model is characterized by fewer particles with higher momentum. By contrast, Pythia 8.145 4C, Pythia DW, and Herwig++ 2.5.1 UE7 have
too few high-$\varz$ particles, with variations again larger for $R$ = 0.6.  
Pythia DW exhibits an excess at mid-$\varz$ at low jet momenta, which is seen at progressively lower $\varz$ values as the jet momentum 
increases, implying an excess of particles with a momentum of roughly $2\,\GeV$ that is not associated with jet structure.

The transverse momentum $\varptrel$ is in fair agreement ($\sim$$20\%$) at low-to-moderate values for all MC generators except Herwig++ 2.4.2.  
At the lowest jet momenta and highest measurable $\varptrel$, Phojet and Herwig++ 2.5.1 UE7 have an excess of particles, while Pythia DW has a deficit.
At higher jet momenta, the data have more high-$\varptrel$ particles than any tune, with Perugia 2010 and Perugia HARD giving the closest description
and Perugia SOFT the furthest.
Perugia 2010 and Perugia HARD agree better than do the other Perugia tunes.

The  AMBT1, Pythia 8.145 4C, and Herwig++ 2.5.1 UE7 tunes provide a good description of the charged particle number density $\rho_{ch}(r)$ 
(Figure~\ref{fig:paper_rdA}) at all radii.
Phojet and the Perugia tunes (especially SOFT) have an excess of particles very close to the jet axis, which is most pronounced at high jet momentum and for $R$ = 0.6; 
Perugia 2010 agrees better in this region than do the other Perugia tunes.  
At high $r$, Phojet and all Pythia tunes except AMBT1 and Perugia SOFT have too few particles.  However, the disagreement is less pronounced than is seen at high $\varptrel$,
implying that high-radius soft particles from the underlying event are better-described than high-radius hard radiation.

\section{Conclusions}

A measurement is presented of the charged particle jet cross section as a function of transverse momentum and rapidity, 
along with the transverse momentum, longitudinal momentum fraction, and number density
as a function of radius for charged particles within these jets, using early $7\,\TeV$ LHC collision data collected with the ATLAS detector.
The study of jets with tracks allows for precise measurements of low-momentum jets and their properties, thus complementing calorimeter-based jet measurements
and allowing the study of the transition from soft collisions to jet production in the perturbative regime of QCD.
It also provides additional observables for consideration in the tuning of MC event generators, which complement existing ``minimum bias'' and underlying event measurements.

No tune or model presented here agrees with all quantities measured within their uncertainties, suggesting that future MC tunes may be improved.
Difficulty in modeling the transition between soft and perturbative physics is indicated by disagreements between data and all MC distributions in 
the 10--$20\,\GeV$ range in the dependence of the charged jet cross section on jet momentum.  
Dependence of the cross section on rapidity is consistent with predictions.
Particles with large transverse momentum $p_{\mathrm{T}}^{rel}$ with respect to the jet that contains them are produced more copiously than any model predicts,
as are jets with large charged particle multiplicity.
The longitudinal momentum fraction $\varz$ is best described by the Pythia 6.421 AMBT1 tune.
The charged particle number density $\rho_{ch}(r)$ is well-described by the Pythia 6.421 AMBT1, Pythia 8.145 4C, and Herwig++ 2.5.1 UE7 tunes.
With the exception of the Herwig++ 2.4.2 default tune, which greatly disagrees with these measurements, 
all models appear to underestimate the contribution of the underlying event required to model the data.

\section{Acknowledgements}

We thank CERN for the very successful operation of the LHC, as well as the
support staff from our institutions without whom ATLAS could not be
operated efficiently.

We acknowledge the support of ANPCyT, Argentina; YerPhI, Armenia; ARC,
Australia; BMWF, Austria; ANAS, Azerbaijan; SSTC, Belarus; CNPq and FAPESP,
Brazil; NSERC, NRC and CFI, Canada; CERN; CONICYT, Chile; CAS, MOST and
NSFC, China; COLCIENCIAS, Colombia; MSMT CR, MPO CR and VSC CR, Czech
Republic; DNRF, DNSRC and Lundbeck Foundation, Denmark; ARTEMIS, European
Union; IN2P3-CNRS, CEA-DSM/IRFU, France; GNAS, Georgia; BMBF, DFG, HGF, MPG
and AvH Foundation, Germany; GSRT, Greece; ISF, MINERVA, GIF, DIP and
Benoziyo Center, Israel; INFN, Italy; MEXT and JSPS, Japan; CNRST, Morocco;
FOM and NWO, Netherlands; RCN, Norway; MNiSW, Poland; GRICES and FCT,
Portugal; MERYS (MECTS), Romania; MES of Russia and ROSATOM, Russian
Federation; JINR; MSTD, Serbia; MSSR, Slovakia; ARRS and MVZT, Slovenia;
DST/NRF, South Africa; MICINN, Spain; SRC and Wallenberg Foundation,
Sweden; SER, SNSF and Cantons of Bern and Geneva, Switzerland; NSC, Taiwan;
TAEK, Turkey; STFC, the Royal Society and Leverhulme Trust, United Kingdom;
DOE and NSF, United States of America.

The crucial computing support from all WLCG partners is acknowledged
gratefully, in particular from CERN and the ATLAS Tier-1 facilities at
TRIUMF (Canada), NDGF (Denmark, Norway, Sweden), CC-IN2P3 (France),
KIT/GridKA (Germany), INFN-CNAF (Italy), NL-T1 (Netherlands), PIC (Spain),
ASGC (Taiwan), RAL (UK) and BNL (USA) and in the Tier-2 facilities
worldwide.

\bibliographystyle{apsrev4-1}
\bibliography{TrackJet_Final}
\onecolumngrid
\clearpage
\input{atlas_authlist_TrackJet}

\end{document}

%% file: atlas_authlist_TrackJet.tex
\begin{flushleft}
{\Large The ATLAS Collaboration}

\bigskip

G.~Aad$^{\rm 48}$,
B.~Abbott$^{\rm 111}$,
J.~Abdallah$^{\rm 11}$,
A.A.~Abdelalim$^{\rm 49}$,
A.~Abdesselam$^{\rm 118}$,
O.~Abdinov$^{\rm 10}$,
B.~Abi$^{\rm 112}$,
M.~Abolins$^{\rm 88}$,
H.~Abramowicz$^{\rm 153}$,
H.~Abreu$^{\rm 115}$
% ...
\bigskip

% ...
$^{10}$ Institute of Physics, Azerbaijan Academy of Sciences, Baku, Azerbaijan\\
$^{11}$ Institut de F\'isica d'Altes Energies and Universitat Aut\`onoma  de Barcelona and ICREA, Barcelona, Spain\\
% ...
$^{48}$ Fakult\"{a}t f\"{u}r Mathematik und Physik, Albert-Ludwigs-Universit\"{a}t, Freiburg i.Br., Germany\\
$^{49}$ Section de Physique, Universit\'e de Gen\`eve, Geneva, Switzerland\\
% ...
$^{88}$ Department of Physics and Astronomy, Michigan State University, East Lansing MI, United States of America\\
% ...
$^{111}$ Homer L. Dodge Department of Physics and Astronomy, University of Oklahoma, Norman OK, United States of America\\
$^{112}$ Department of Physics, Oklahoma State University, Stillwater OK, United States of America\\
% ...
$^{115}$ LAL, Univ. Paris-Sud and CNRS/IN2P3, Orsay, France\\
% ...
$^{118}$ Department of Physics, Oxford University, Oxford, United Kingdom\\
% ...
$^{153}$ Raymond and Beverly Sackler School of Physics and Astronomy, Tel Aviv University, Tel Aviv, Israel\\
% ...
\end{flushleft}
